\documentclass{aa}

\usepackage[varg]{txfonts}
\usepackage{graphicx}
\usepackage{amsmath}
\usepackage[colorlinks=true,linkcolor=magenta,citecolor=blue,filecolor=cyan,urlcolor=black,final=true]{hyperref}
\usepackage[normalem]{ulem} 
\usepackage{xcolor} 

\makeatletter
\renewcommand*\aa@pageof{, page \thepage{} of \pageref*{LastPage}}
\makeatother
 
\begin{document} 

   \title{Electron acceleration and radio emission following the early interaction of two coronal mass ejections}
	\titlerunning{Electron acceleration and radio emission resulting from two interacting coronal mass ejections}

   \author{D.~E.~Morosan \inst{1}
        \and
        E.~Palmerio \inst{1,2} 
        \and
        J.~E.~R{\"a}s{\"a}nen \inst{1}
        \and
        E.~K.~J.~Kilpua \inst{1} 
        \and
        J.~Magdaleni{\'c} \inst{3}
        \and
        B.~J.~Lynch \inst{2}
          \and
        A.~Kumari \inst{1}
        \and
        J.~Pomoell \inst{1}
                  \and
        M.~Palmroth \inst{1,4}}

   \institute{Department of Physics, University of Helsinki, P.O. Box 64, FI-00014 Helsinki, Finland \\
              \email{diana.morosan@helsinki.fi}
         \and
            Space Sciences Laboratory, University of California--Berkeley, Berkeley, CA 94720, USA
         \and
             Solar--Terrestrial Centre of Excellence---SIDC, Royal Observatory of Belgium, 1180 Brussels, Belgium
        \and
            Space and Earth Observation Centre, Finnish Meteorological Institute, P.O. Box 503, FI-00014 Helsinki, Finland
             }

   \date{Received ; accepted }

 
  \abstract
    {Coronal mass ejections (CMEs) are large eruptions of magnetised plasma from the Sun that are often accompanied by solar radio bursts produced by accelerated electrons.}
    {A powerful source for accelerating electron beams are CME-driven shocks, however, there are other mechanisms capable of accelerating electrons during a CME eruption. So far, studies have relied on the traditional classification of solar radio bursts into five groups (Type I--V) based mainly on their shapes and characteristics in dynamic spectra. Here, we aim to determine the origin of moving radio bursts associated with a CME that do not fit into the present classification of the solar radio emission.}
    {By using radio imaging from the Nan{\c c}ay Radioheliograph, combined with observations from the Solar Dynamics Observatory, Solar and Heliospheric Observatory, and Solar Terrestrial Relations Observatory spacecraft, we investigate the moving radio bursts accompanying two subsequent CMEs on 22 May 2013. We use three-dimensional reconstructions of the two associated CME eruptions to show the possible origin of the observed radio emission.}
    {We identified three moving radio bursts at unusually high altitudes in the corona that are located at the northern CME flank and move outwards synchronously with the CME. The radio bursts correspond to fine-structured emission in dynamic spectra with durations of ${\sim}1$~s, and they may show forward or reverse frequency drifts. Since the CME expands closely following an earlier CME, a low coronal CME--CME interaction is likely responsible for the observed radio emission.}
    {For the first time, we report the existence of new types of short duration bursts, which are signatures of electron beams accelerated at the CME flank. Two subsequent CMEs originating from the same region and propagating in similar directions provide a complex configuration of the ambient magnetic field and favourable conditions for the creation of collapsing magnetic traps. These traps are formed if a CME-driven wave, such as a shock wave, is likely to intersect surrounding magnetic field lines twice. Electrons will thus be further accelerated at the mirror points created at these intersections and eventually escape to produce bursts of plasma emission with forward and reverse drifts.}


\maketitle


\section{Introduction}

{One of the most prominent sources of particle acceleration in our Solar System are large eruptions of magnetised plasma from the Sun called coronal mass ejections (CMEs). When they  propagate faster than the characteristic speed of the ambient medium, they drive shock waves where electrons can be accelerated to high energies. In addition, the associated magnetic reconfiguration processes during the eruption can lead to the acceleration of particles. The accelerated particles can in turn generate emission at radio wavelengths through the plasma emission mechanism \citep[e.g.][]{kl02}.}

{Shocks driven by CMEs can be observed in white light as large, extended structures of fainter emission surrounding the brighter CME bubble \citep{vo03,vo13}. CME shocks are usually much larger than the underlying magnetic flux rope \citep[e.g.][]{vo13}, consisting of a bubble-like structure surrounding the flux rope and CME ejecta \citep{kw14, liu19}. However, other regions are associated with CMEs that can accelerate electrons, such as reconnecting current sheets forming at the wake of CMEs \citep{ka92} or magnetic traps that formed due to complex magnetic field configurations following CME eruptions \citep{ma02}. Collapsing magnetic traps, in particular, have been previously suggested as a mechanism to further accelerate protons and electrons due to the expansion of the CME shock into the surrounding coronal field lines \citep{ma02,po08,ko12}. Strong particle acceleration has also been observed in extreme space weather events as a result of CME--CME interactions at larger heights in the corona \citep{di13,go16}, however it is unclear if these types of interactions cause the enhancement of accelerated particles \citep{ri03}}. Some of these events are also accompanied by significant radio emission, such as interplanetary radio bursts \citep{go16, liu17, pa19, ma12}.

{The most obvious manifestations of shocks at radio wavelengths are a class of radio bursts called Type II bursts, which consist of slowly drifting emission lanes observed in dynamic spectra at the fundamental and harmonic of the plasma frequency \citep{ma96,ne85}. Type II emission sources are closely associated with expanding CME-driven shocks in the corona \citep{st74, ma12, liu09, zu18}. Type II bursts can often show split-band lanes \citep{vr01,vr02} and numerous fine structures composing the emission lanes \citep{ma20}. Fine-structured bursts called `herringbones', which can accompany Type II bursts or occur on their own \citep{ho83,ca87,ca89}, also represent individual electron beams accelerated by a CME shock \citep{ca13,mo19a}. Type IIs and herringbones can be used to study the shock kinematics either through their drift rate observed in dynamic spectra, through radio images, or through multi-point observations, since the emission sources are expected to move in the direction of the CME expansion \citep[e.g.][]{re07, liu13, ma14, mo19a, zh19, maguire20}.}

{However, there are other radio bursts accompanying CMEs that can also show propagation associated with the CME expansion, namely moving Type IV radio bursts \citep{bo57}. Type IV bursts are a broadband continuum emission, often with superimposed fine structures \citep[e.g.][]{ma06, bo16}. These bursts are commonly believed to be generated either by electrons trapped inside the CME flux rope emitting gyro-synchrotron radiation \citep{du73,ba01}, electrons accelerated at a reconnecting current sheet forming at the wake of an erupting flux rope, thereby generating bursts of plasma emission \citep{ka92,vr03,mo19b}, or emission at the plasma frequency from inside the CME \citep{st78, va19}. Recent observations have also shown moving Type IV sources that are closely related with the propagation of CME flanks \citep{mo20b} and Type II-like lanes have also been identified inside a Type IV continuum emission \citep{chr18}. It is therefore feasible that numerous other signatures associated with CMEs that are caused by fast electron beams have been overlooked as a result of adhering to the classical categorisation scheme of solar radio bursts.}

{The availability of observations from the  Solar Terrestrial Relations Observatory \citep[STEREO;][]{ka08} in recent times has brought a third dimension to the studies of eruptive events and has allowed us to better understand the relationship between CMEs and associated radio emission \citep{ma19, mo19a, mo20a, chr20}. Significant plane-of-sky projection effects of radio sources have been identified, even in observations of CMEs originating on the solar limb. Therefore, a three-dimensional picture of the radio emission relative to the CME is necessary to determine the locations of electron acceleration regions. At larger distances, triangulation techniques have so far been used to determine the location of Type II radio sources \citep[e.g.][]{ma12}. So far, the 3D location of electron acceleration regions in relation to the propagating CME structure and ambient corona early during the eruption is poorly known, largely due to limitations of the previously available observational capabilities. }  

{In this paper, we present the first report of moving radio bursts observed at unusually high altitudes (1.5--2.2\,$R_\odot$ at 150~MHz) and associated with two subsequent CMEs that erupted on 22~May~2013. These bursts do not fit in either the Type II nor Type IV classifications. Using a combination of multi-spacecraft observations and 3D reconstructions we are able to show that the moving bursts originate at a location where the CME propagates in the wake of an earlier CME, presumably creating conditions where electrons can be accelerated effectively in contracting magnetic bottles. In Sect.~\ref{sec:analysis}, we give an overview of the observations and employed data analysis techniques. In Sect.~\ref{sec:results}, we present the results, which are further discussed in Sect.~\ref{sec:discussion}. The conclusions are presented in Sect.~\ref{sec:conclusion}.}


\section{Observations and data analysis} \label{sec:analysis}

\subsection{Radio emission}

\begin{figure*}[ht]
    \centering
    \includegraphics[width=0.8\linewidth]{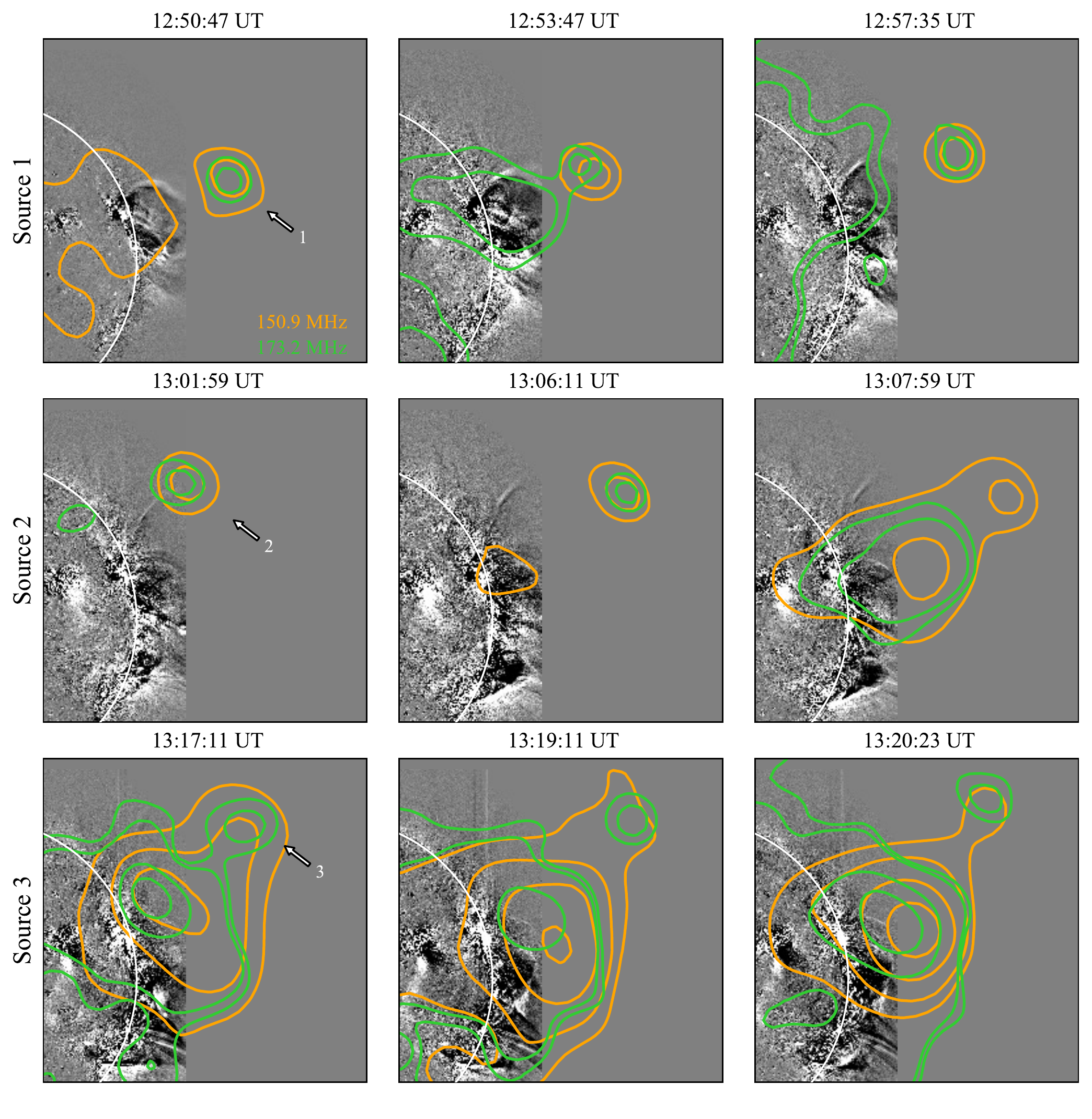}
    \caption{Moving radio bursts associated with the expansion of the 22~May~2013 CME in the solar corona. NRH radio contours at 150.9~MHz (orange) and 173.2~MHz (green) are overlaid on AIA 211~{\AA} running-difference images in each panel. Three moving radio bursts were identified and labelled as Source 1 (top panels), Source 2 (middle panels), and Source 3 (bottom panels). The AIA images show the evolution of the CME through time. }
    \label{fig:fig1}
\end{figure*}

\begin{figure*}[ht]
\centering
    \includegraphics[width=0.8\linewidth]{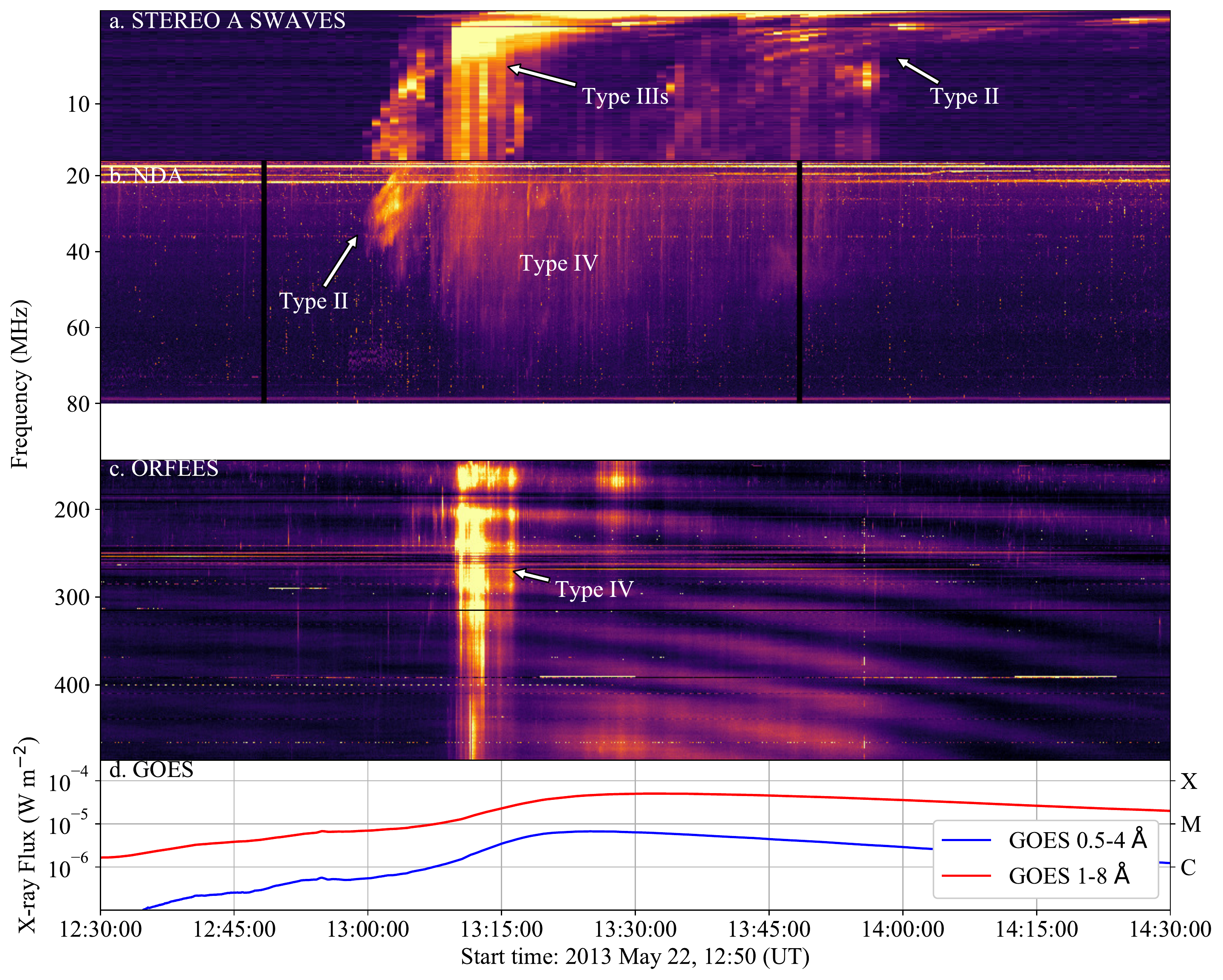}
    \caption{Dynamic spectrum of the radio bursts associated with the 22~May~2013 CME and solar X-ray flux. (a), (b), (c) Composite dynamic spectrum from SWAVES-A, NDA, and ORFEES showing the radio activity associated with the CME in the following frequency ranges: 1--16, 16--80, and 144--484~MHz, respectively. (d) GOES solar X-ray flux indicating the occurrence of an M5.0 solar flare.}
    \label{fig:fig2}
\end{figure*}

{On 22 May 2013, three moving radio sources were observed in images from the Nan{\c c}ay Radioheliograph \citep[NRH;][]{ke97}. The moving radio sources are labelled in order of their appearance in Fig.~\ref{fig:fig1}: Source 1 (top panels), Source 2 (middle panels), and Source 3 (bottom panels). These bursts are shown as radio intensity contours at 150 (orange) and 173~MHz (green), overlaid on running-difference images at 211~{\AA} from the Atmospheric Imaging Assembly \cite[AIA;][]{le12} onboard the Solar Dynamics Observatory \citep[SDO;][]{pe12}. Few other stationary radio sources were observed at the same time but at different locations, closer to the active region (see e.g.\ the other radio contours in Fig.~\ref{fig:fig1} that are not labelled by arrows).}

{In the combined dynamic spectra from the Radio and Plasma Wave Investigation \citep[SWAVES;][]{bougeret2008} onboard STEREO, the Nan{\c c}ay Decametric Array \citep[NDA;][]{bo80}, and the ORFEES (Observation Radio pour FEDOME et l'{\'E}tude des {\'E}ruptions Solaires) radio spectrograph, an intense radio event was observed over a wide frequency band. The radio emission started at ${\sim}$13:00~UT and lasted for a few hours (Fig.~\ref{fig:fig2}a--c).}

{At frequencies below 100~MHz, numerous Type II lanes were observed (Fig.~\ref{fig:fig2}a--b), indicating the presence of a shock wave. The kinematics of these Type II lanes in relation to the associated CME (presented in the following section) were studied in \citet{ma16} and \citet{pa19}. In the same frequency range, a prominent Type IV burst was observed, lasting for up to 1~hour (Fig.~\ref{fig:fig2}b). Drifting structures, possibly other Type II lanes, were superimposed on the Type IV continuum. At frequencies above 140~MHz, a short duration ($\sim$30-minutes), broadband continuum emission was observed starting from $\sim$13:10~UT (labelled as a Type IV in Fig.~\ref{fig:fig2}c). Shortly after the first Type II lanes, a group of Type III bursts was observed in the STEREO/SWAVES frequency range. The preceding radio emission (ORFEES dynamic spectrum in Fig.~\ref{fig:fig2}c) is dominated by fine-structured bursts, most of which are Type I bursts constituting a noise storm. The noise storm originated from an unrelated active region located on the eastern solar hemisphere. The majority of the radio emission was observed during the rise time of an M5.0-class flare (the peak of the X-ray flux occurred at 13:32~UT in Fig.~\ref{fig:fig2}d from the Geostationary Environmental Satellites; GOES). The most intense Type II lanes were also observed during the rise time of the flare, which indicates the presence of an early shock propagating in the low corona. However, both the Type II and Type IV emissions continue during the decay phase of the flare, with the Type II lanes in particular observed for longer than the low-frequency Type IV burst. All these aspects indicate a long-lasting electron acceleration process as the CME propagates further out into interplanetary space.}

\subsection{Eruptive event}

\begin{figure*}[ht]
\centering
    \includegraphics[width=0.9\linewidth]{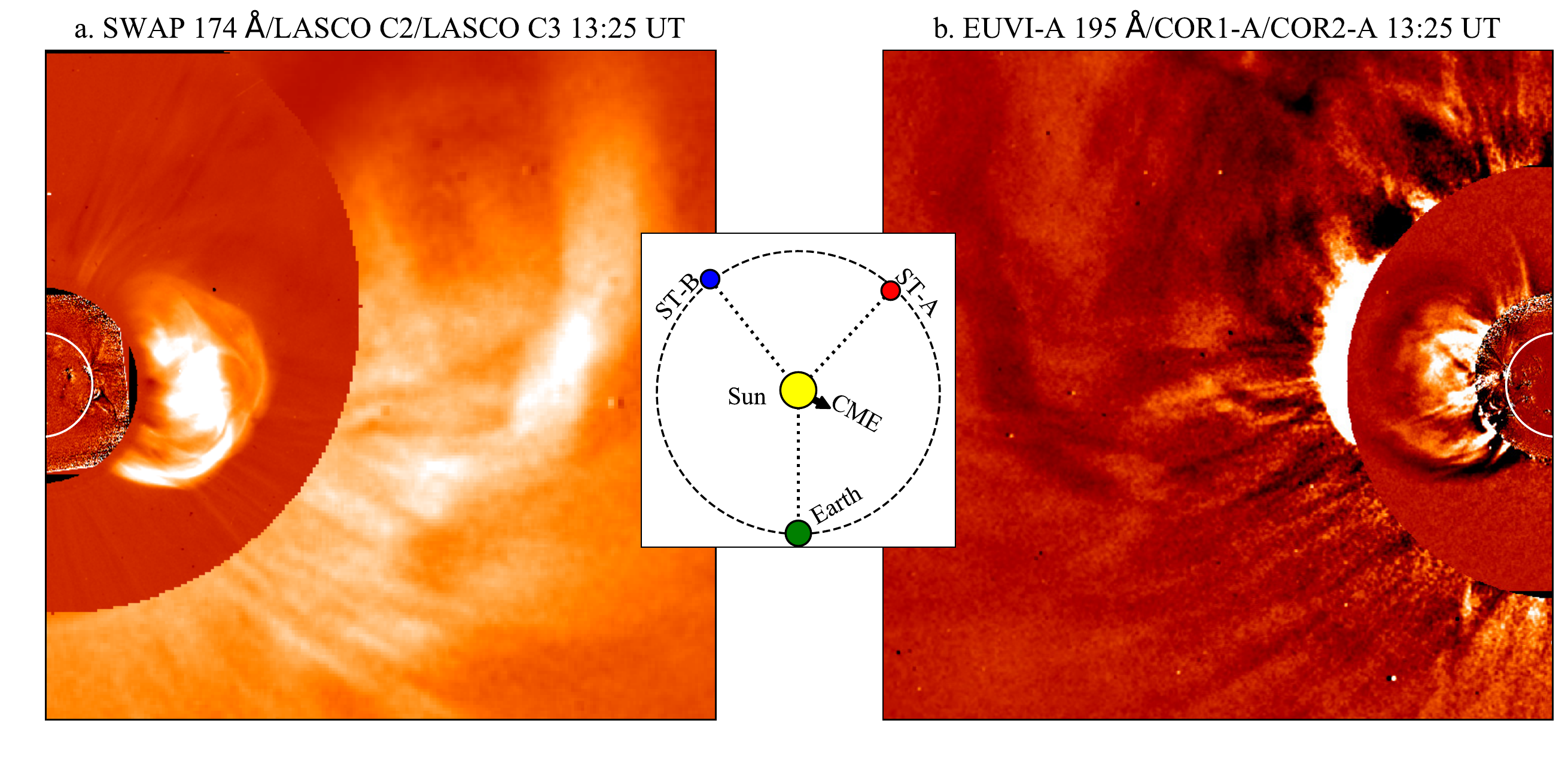}
    \caption{The coronal mass ejections observed on 22~May~2013 from Earth and STEREO-A. (a) The CMEs observed on 22 May 2013 from Earth's perspective in a composite image from SWAP 174~{\AA}, LASCO/C2, and LASCO/C3. The CME associated with the radio emission in Fig.~\ref{fig:fig2} can be seen in the LASCO/C2 field of view (inner), while the earlier CME has expanded higher up in the corona and can be seen in the LASCO/C3 field of view (outer). (b) The CMEs observed on 22 May 2013 from STEREO-A's perspective in a composite image from EUVI-A 195~{\AA}, COR1-A, and COR2-A. The main CME is visible in the COR1-A field of view (inner), while the earlier CME is visible in the COR2-A field of view (outer). The middle inset in the figure shows the propagation direction of the CME and the location of Earth and the STEREO spacecraft with respect to the Sun. }
    \label{fig:fig3}
\end{figure*}

{The AIA images with overlaid NRH radio sources (Fig.~\ref{fig:fig1}) show that the moving radio bursts occur concurrently with a CME. The radio sources appear to follow the propagation direction of the CME which accompanies the GOES M5.0-class flare (Fig.~\ref{fig:fig2}d). The NRH moving radio bursts seem to be associated with the northern CME flank (for a full evolution of the moving radio bursts and the CME, see Movie~1 accompanying this paper). }

{The CME was well observed in remote-sensing data from instruments onboard the Solar and Heliospheric Observatory \citep[SOHO;][]{do95}, the Project for On Board Autonomy 2 \citep[PROBA2;][]{sa13}, and SDO that are  located near Earth. The CME was also observed by the twin STEREO spacecraft, separated from Earth by $137^{\circ}$ and $141^{\circ}$, respectively. Figure~\ref{fig:fig3} shows the CME as it propagates through the solar corona from the perspectives of Earth (Fig.~\ref{fig:fig3}a) and STEREO-A (Fig.~\ref{fig:fig3}b). The first observation of the CME in white-light coronagraph images was at 12:55~UT by STEREO-A's inner coronagraph COR1, which is part of the Sun Earth Connection Coronal and Heliospheric Investigation \citep[SECCHI;][]{ho08} suite. From Earth's perspective, the CME was observed first at 13:25~UT with The Large Angle Spectrographic Coronagraph \citep[LASCO;][]{br95} C2 instrument onboard SOHO.}

{Sun-monitoring spacecraft in Earth's orbit, such as SOHO and SDO, were located at ideal locations from STEREO-A such that the CME is well observed from different sides. SOHO and SDO are separated at this time by $137^{\circ}$ from STEREO-A, with their mid-point at $68.5^{\circ}$ (see the inset in Fig.~\ref{fig:fig3}). The CME longitude deduced from observations is ${+}67^{\circ}$ \citep{pa19} in Stonyhurst coordinates. Therefore, the locations of the spacecraft close to Earth and STEREO-A are almost perfectly symmetrical with respect to the CME propagation direction (Fig.~\ref{fig:fig3}). The CME is shown from Earth's perspective in a combined image at $\sim$13:25~UT from the Sun-Watcher with Active Pixel System and Image Processing \citep[SWAP;][]{se13} onboard PROBA2, LASCO/C2, and LASCO/C3 (Fig.~\ref{fig:fig3}a). From the perspective of STEREO-A, the CME is shown in a combined image at $\sim$13:25~UT from STEREO-A's Extreme Ultraviolet Imager (EUVI), inner coronagraph COR1, and outer coronagraph COR2 (Fig.~\ref{fig:fig3}b).}

{Careful investigation of images from LASCO's outer coronagraph C3 and STEREO-A's outer coronagraph COR2 (Fig.~\ref{fig:fig3}) shows that the CME did not expand through the undisturbed solar corona, but behind an ongoing earlier CME. The full evolution of the main CME in the low corona observed by SWAP and both CMEs in the high corona observed by COR2-A can be seen in Movies 2 and 3 accompanying this paper, respectively. The preceding CME was first observed in white-light images from COR1-A at 09:24~UT, occurring ${\sim}4$~hours before the main CME studied here. Therefore, the main CME, which occurred at ${\sim}$13:00~UT, propagated through a strongly disturbed coronal plasma consisting of material and magnetic field lines behind the earlier CME. The main CME propagated through the outer corona with a de-projected speed of ${\sim}1500$~km/s, as reported by \citet{pa19}, and eventually caught up with the preceding CME. LASCO/C3 observations show the CMEs merging at an altitude of ${\sim}20$\,$R_{\odot}$ \citep[reported by][based on plane-of-sky images]{pa19}. The resulting ejecta was observed in situ at STEREO-A, while the interplanetary shock driven by the merged structure was detected at both Earth and STEREO-A \citep{pa19}.}


\section{Results} \label{sec:results}

\begin{figure}[ht]
\centering
    \includegraphics[width=0.95\linewidth]{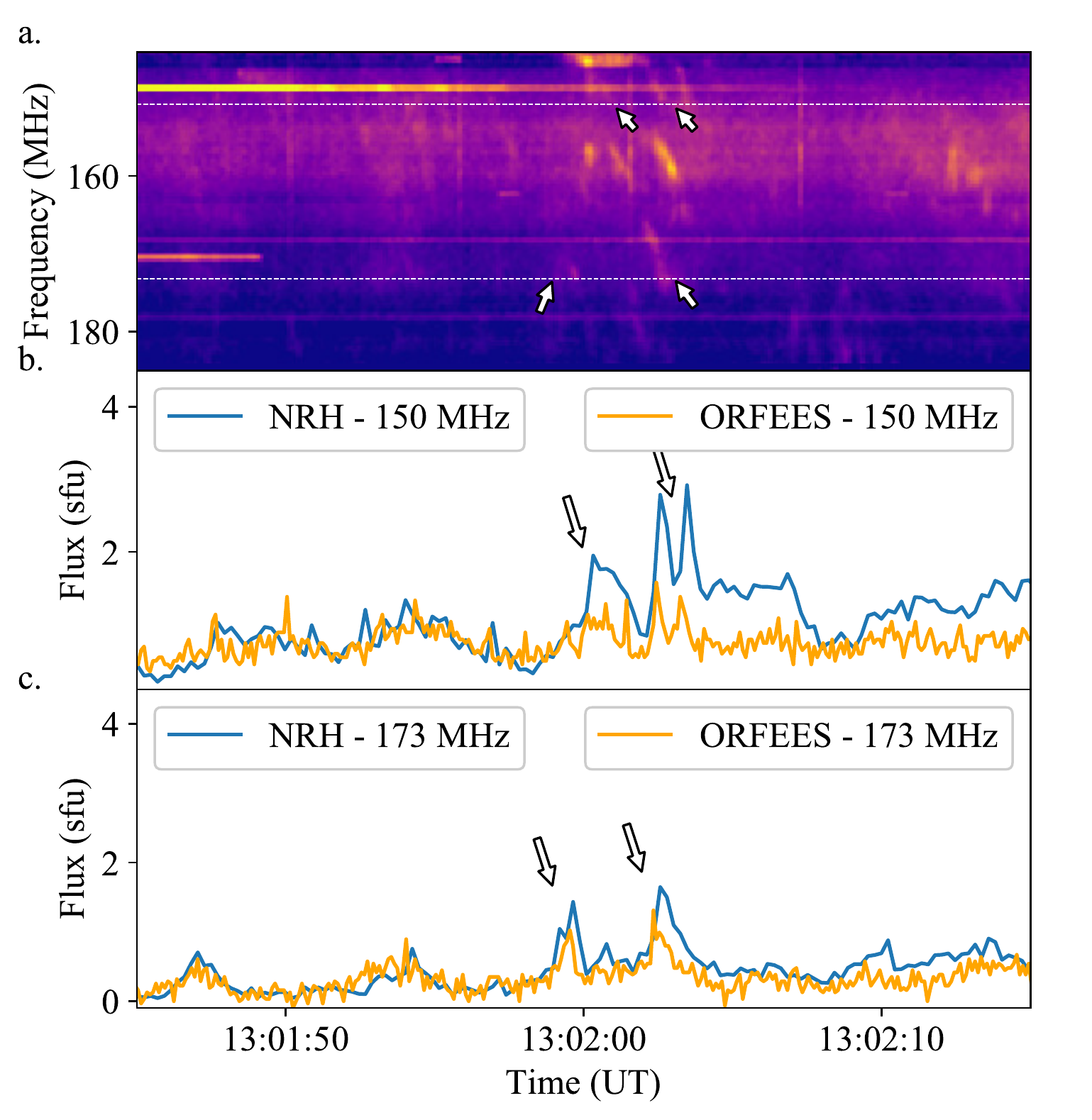}
    \caption{Spectral features corresponding to the moving radio burst labelled as Source 2 in Fig.~\ref{fig:fig1}. Figures corresponding to Sources 1, 2, and 3 are included in Appendix~\ref{app:a}. (a) Zoomed-in view of the ORFEES dynamic spectrum showing a 30-s period starting at 13:01:45~UT. The first arrow points at one of the Source 2 bursts shown in the middle--left panel of Fig.~\ref{fig:fig1}. (b--c) Time series of the NRH flux density of Source 2 in sfu and ORFEES normalised intensity in arbitrary units at (b) 150 and (c) 173~MHz.}
    \label{fig:fig4}
\end{figure}

{The moving radio bursts presented in Fig.~\ref{fig:fig1} are observed at the two lowest NRH frequencies, 150 and 173~MHz, and do not show any circular polarisation. In order to identify the possible emission mechanism, origin of the energetic electrons generating this emission, and their relation to the CME we examine the spectral characteristics, kinematics, and 3D location of these moving bursts.}

\subsection{Spectral characteristics of the moving radio bursts}

{A detailed dynamic spectrum shows Source 2 for a period of 30~s starting from 13:01:45~UT (Fig.~\ref{fig:fig4}a). Similar figures, corresponding to Sources 1 and 3, together with an additional example for Source 2, are shown in Appendix~\ref{app:a}. The Type I noise storm is the dominant emission in the ORFEES dynamic spectra shown in Fig.~\ref{fig:fig4}a and in the figures in Appendix~\ref{app:a}. However, at times when the moving radio bursts appear in NRH images, they are sometimes brighter than the dominating noise storm. The moving sources correspond to individual bursts that are separate from the noise storm in the ORFEES dynamic spectra.}

{To verify the association between the features observed in ORFEES spectra and the moving radio sources, we compute the flux densities of the moving bursts and compare them to the corresponding ORFEES time series. The flux densities of the moving radio bursts observed by NRH are estimated at 150 and 173~MHz for all three sources. The flux densities are estimated inside a zoomed-in box covering the full extent and movement of the radio source, and they include the pixels with levels >20\% of the maximum intensity levels in each box. We choose a threshold of 20\% to include the full extent of the radio source and exclude quiet-Sun or other types of weak emission. The flux densities are estimated in solar flux units (sfu; where 1~sfu = $10^{-22}$~W~m$^{-2}$~Hz$^{-1}$) for total intensity (Stokes I). The NRH flux densities (blue time series in Fig.~\ref{fig:fig4}b--c at 150 and 173~MHz) show peaks of bursty emission. Similar behaviour is shown in the ORFEES normalised intensities at 150 and 173~MHz (orange time series in Fig.~\ref{fig:fig4}b--c) that are extracted from the dynamic spectrum in Fig.~\ref{fig:fig4}a.  }

\begin{figure*}[ht]
\centering
    \includegraphics[width=0.95\linewidth]{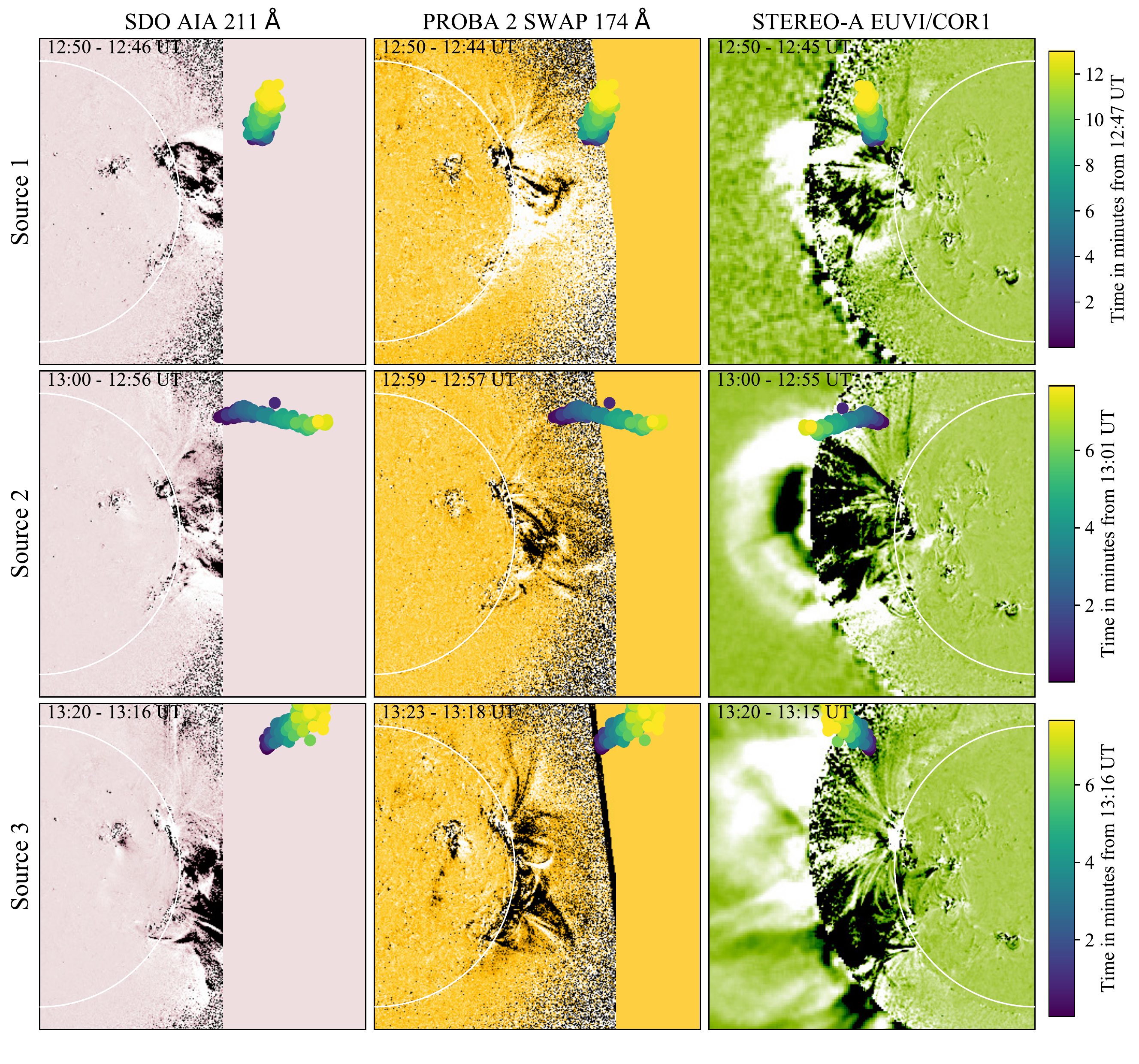}
    \caption{Evolution of the centroids of the moving radio sources through time. The centroids of the three moving sources are shown from top to bottom overlaid on SDO/AIA (left), PROBA2/SWAP (middle), and STEREO/EUVI-A and COR1-A (right) running-difference images of the Sun. The colour scale for the centroids represents minutes from the start time of each moving radio burst. The bursts are located at the northern CME flank in both the Earth (SDO and PROBA2) and STEREO-A perspectives. }
    \label{fig:fig5}
\end{figure*}

{The bursts marked by the white arrows in the top panel of Fig.~\ref{fig:fig4} correspond to the moving radio burst that we labelled as Source 2. In particular, the burst in Fig.~\ref{fig:fig4} marked by the leftmost white arrow corresponds to the NRH burst in the left middle panel of Fig.~\ref{fig:fig1}.  We note that there is a slight delay of ${\sim}0.5$~s between the NRH and ORFEES time stamps. We also note that ORFEES is less sensitive than the NRH, which may be the reason why it does not observe the moving radio sources when they are fainter compared to the examples shown in Fig.~\ref{fig:fig4} and Appendix~\ref{app:a}.  The majority of bursts identified in the ORFEES spectrum that correspond to the moving radio bursts (including those presented in the figures in Appendix~\ref{app:a}) represent fine-structured, narrow-band bursts. These bursts have forward or reverse drifts and durations of ${\sim}1$~s. Thus, these drifting bursts resemble herringbones. However, they occur on their own instead of occurring in large clusters and they also do not show the presence of a `backbone' with forward and reverse drift bursts on either side \citep{mo19a}. Similarly to what was recently reported by \citet{ma20}, other bursts somewhat resemble the fine structures composing Type II bursts (see Fig.~\ref{fig:figa2}). The morphological resemblance of drifting bursts and herringbones indicates that also the emission mechanism of both types of bursts could be the same, that is, plasma emission. In such a scenario, the accelerated electron beams, associated with drifting bursts, would travel in various directions, similar to the way herringbones are generated \citep{ho83, ca87}. However, we note that there are also other potential sources to these electron beams than acceleration at the CME shock (see the Introduction). The knowledge on the spatial location of these fine-structured bursts will provide some additional insight into the origin of these bursty electron beams, which is further investigated in the following sections.}


\subsection{Moving radio sources and CME kinematics}

\begin{figure*}[ht]
\centering
    \includegraphics[width=0.9\linewidth, trim = {0px 0px 0px 0px}, clip]{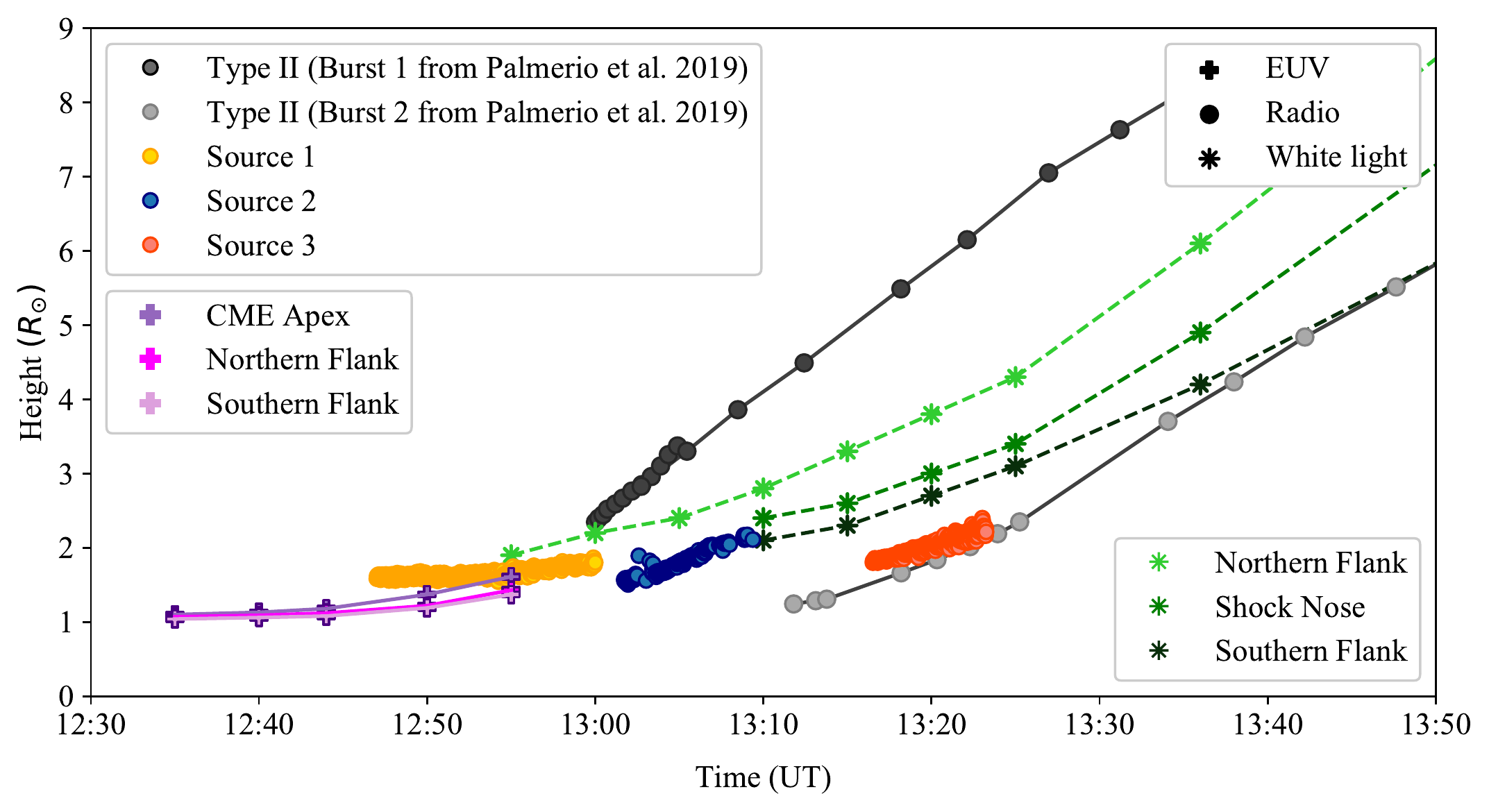}
    \caption{Kinematics of the moving radio bursts and Type II lanes in relation to the CME expansion. The Type II bursts and CME kinematics were previously studied by \citet{pa19}. The figure is adapted from \citet{pa19}, to which we included the three moving radio sources studied here. The EUV measurements are taken with PROBA2/SWAP and STEREO/EUVI-A, the white-light ones with LASCO/C2--C3 and STEREO/COR1--COR2-A, and the radio ones with NDA and the WAVES instruments onboard Wind \citep{og97} and both STEREO spacecraft.}
    \label{fig:fig6}
\end{figure*}

{The moving radio sources are observed at unusually high altitudes (up to 2.2\,$R_\odot$ in Fig.~\ref{fig:fig1}) for emission at 150 and 173~MHz. In the background solar corona, a plasma frequency level of 150.9~MHz corresponds to a height range of 1.05--1.35\,$R_\odot$ for fundamental emission and 1.20--1.65\,$R_\odot$ for harmonic emission, based on electron density models of the corona \citep{new61,sa77}. The centroids of the radio emission at 150.9~MHz show outward movement in the direction of the CME expansion (Fig.~\ref{fig:fig5}). Source 1 and Source 3 move in a north-westerly direction, along with the CME flank, while Source 2 moves in a westerly direction with the CME apex. However, from Earth's viewpoint, this plane-of-sky movement can be affected by line-of-sight projection effects. While a significant portion of this movement occurs close to the plane of sky, the sources could also be directed at an angle towards or away from the observer. We thus aim to reconstruct a 3D perspective using CME observations from STEREO-A that was at an ideal location to provide a view of the eruption from the other side (see the inset in Fig.~\ref{fig:fig4}a). STEREO-A provides observations in both extreme ultra-violet (EUV) and white light, extending to larger heights than SDO or PROBA2 (the STEREO/COR1 field of view extends between 1.5 and 4\,$R_{\odot}$).}

{To determine the position of the radio sources in relation to the two CMEs in the STEREO-A perspective, we de-project the radio source centroids from their plane-of-sky view. Assuming that the radio bursts are emitted close to the plane of sky (i.e.\ the \textit{z}-coordinate of this emission is ${\sim}0$), it is possible to estimate their approximate location in the STEREO-A perspective. The \textit{z}-coordinates, combined with the plane-of-sky coordinates of the NRH centroids, are then used to project the coordinates of the radio sources onto the STEREO-A plane and to also represent them in 3D. The assumption that the \textit{z}-coordinate of this emission is ${\sim}0$ can be used since the centroids show significant movement in the plane of sky. However, we note that the radio sources could also move towards or away from the observer, from Earth's perspective, and only cross the plane of sky as they propagate. Moreover, the centroids could also be located away from the plane of sky. However, if we consider a large and negative \textit{z}-coordinate, then the centroids would be located too far away from the CME flank in the STEREO-A perspective. If we consider a large and positive \textit{z}-coordinate, then the centroids would be located farther out, close to the CME apex and at even higher altitudes, which is unrealistic for plasma emission. The \textit{z}-coordinate of the radio centroids most likely has values in the range $\pm$300\arcsec, which we assign as the uncertainty range for our de-projection method. This uncertainty range is also consistent with the 3D location of the main CME. Therefore, the assumption that the radio sources propagate close to the plane of sky is a good approximation to determine their location in other perspectives, and relative to the two associated CMEs.}

{The radio centroids of all three sources (coloured with progressing time in Fig.~\ref{fig:fig5}), show the propagation of the radio bursts. The radio sources are superposed on SDO/AIA 211~{\AA} images (left), PROBA2/SWAP 174~{\AA} images (middle), and STEREO-A combined EUVI (at 195~{\AA}) and COR1 images (right). Both Earth (AIA and SWAP) and STEREO perspectives show that the radio sources are located close to the northern CME flank. At the northern flank, the bent field lines due to the passage of the previous CME can be observed in EUV and white-light images (see Figs.~\ref{fig:fig3} and \ref{fig:fig5}). Despite the uncertainty associated with our de-projection technique, the used method would not affect the kinematics of the moving radio bursts in the STEREO perspectives, and these radio bursts would still be associated with the northern CME flank. The unusually high altitudes of these radio bursts may be related to the fact that the CME expands through a strongly disturbed corona.}

\begin{figure*}[ht]
\centering
    \includegraphics[width=9cm, angle = -90, trim = {60px 15px 140px 0px}]{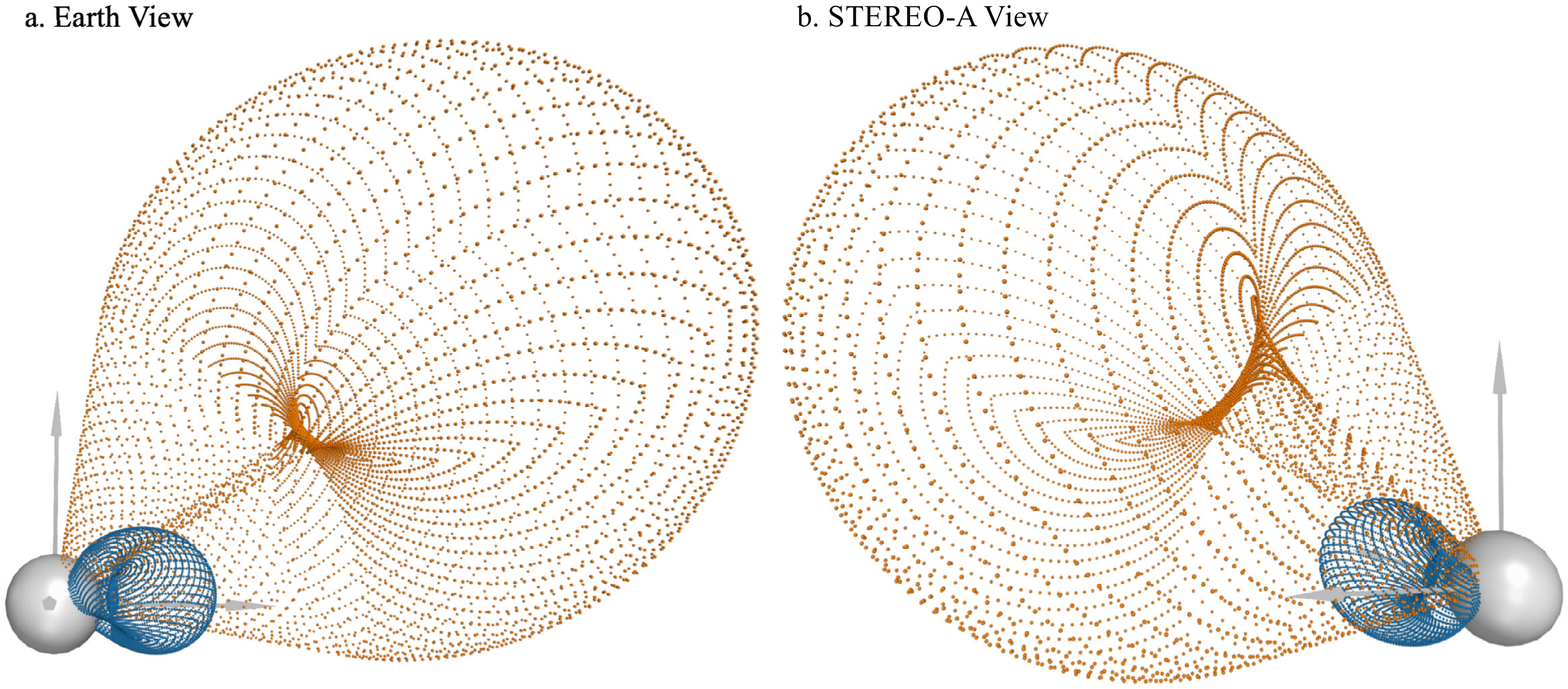}
    \caption{CME expansion in 3D relative to the earlier ongoing CME. The main CME is shown as a wireframe consisting of blue dots, while the ealier CME is shown as an orange wireframe. Two viewpoints are shown from (a) Earth's perspective and (b) STEREO-A's perspective.}
    \label{fig:fig7}
\end{figure*}
 
{The plane-of-sky centroid locations can be compared to the CME kinematics. The CME kinematics have already been studied in relation to the Type II lanes observed at lower frequencies by \citet{pa19}. Unfortunately, imaging at lower frequencies is unavailable for this event and an electron densixty model was used to extract the heights of the Type II lanes observed by NDA and STEREO S-WAVES shown in Fig.~\ref{fig:fig2}. The kinematics of the CME in relation to the Type II lanes is shown in Fig.~\ref{fig:fig6}, adapted from \citet{pa19}, including the additional three moving radio bursts analysed in this study. The early Type II (labelled as Burst 1 in \citealt{pa19}) is faster than the associated CME, while the later Type II (labelled as Burst 2 in \citealt{pa19}) appears to match the southern shock flank kinematics deduced from white light. The plane-of-sky centroids of the moving radio bursts are shown in yellow (Source 1), blue (Source 2), and red (Source 3). Source 2 appears to match the white-light shock nose propagation despite being located at the northern CME flank. This, however, agrees with the propagation direction of the Source 2 centroids in Fig.~\ref{fig:fig5}, that appear to travel radially outwards in the direction of the CME apex, and not laterally outwards with the CME flank. On the other hand, Source 3 has a similar slope to the northern flank shock expansion, which again agrees with the propagation direction of the centroids in Fig.~\ref{fig:fig5}. Source 1 cannot be as clearly related either to the CME flanks or CME apex, since both the apex and CME flanks have similar propagation speeds in the early phase of the eruption, however, as shown in Fig.~\ref{fig:fig5}, it has a similar propagation direction to Source 3.} The plane-of-sky speeds of these sources are $\sim$400~km/s, $\sim$1500~km/s, and $\sim$950~km/s for Sources 1, 2, and 3, respectively. We note that Sources 2 and 3 are particularly fast, and have speeds similar to those of Type II bursts \citep{ma04,pa19}. Source 1 appears to propagate much slower in the plane of sky, however, we note that the CME speed obtained from plane-of-sky EUV observations was also slow during its early stages (see the EUV measurements in Fig.~\ref{fig:fig6}). It should be also noted that estimated plane-of-sky speeds represent a lower limit since it is possible that these sources also propagate towards or away from the observer, therefore all radio sources could be propagating faster and have steeper slopes than those shown in Fig.~\ref{fig:fig6}.


\subsection{3D reconstruction of the locations of electron acceleration}

{The three perspectives provided by SOHO and the two STEREO spacecraft allow us to reconstruct the 3D structure of the CME evolving through the solar corona. We use the Graduated Cylindrical Shell \citep[GCS;][]{the06,the09} model, which consists of a parameterised croissant-shaped shell, to manually trace the CME through time in nearly-simultaneous coronagraph images. We can then model the observed CME in 3D by overlaying the parametrised shell onto the three planes of sky (SOHO, STEREO-A, and STEREO-B). We perform reconstructions from 13:05 to 13:30~UT (at the time of STEREO images) at five-minute intervals during the early propagation of the CME. Since the CME appeared in the LASCO/C2 field of view at 13:25~UT, we perform reconstructions using only the COR1-A and -B viewpoints from 13:05 to 13:20~UT that are guided by the CME parameters obtained from a later time when all three viewpoints are available. Furthermore, we also perform reconstructions of the earlier CME around the same times. The CME reconstruction includes only the CME cavity, excluding the surrounding white light shock.}

\begin{table}
\centering
\begin{tabular}{ |l |r r r r| }

 \hline
                  & Long & Lat & Tilt & Speed\\
 \hline
    Earlier CME   & $73^{\circ}$  & $32^{\circ}$ & $-48^{\circ}$ & 663~km/s\\
    Main CME      & $67^{\circ}$  & $7^{\circ}$ & $-28^{\circ}$ & 1160~km/s \\
 \hline
\end{tabular}
\bigskip
\caption{CME parameters derived from the GCS reconstructions. Longitudes and latitudes are provided in Stonyhurst coordinates, and the tilt is defined positive for counterclockwise rotations.}
\label{tab:tab1}
\end{table}

\begin{figure*}[ht]
\centering
    \includegraphics[width=0.87\linewidth, angle = -90, trim = {10px 110px 30px 10px}]{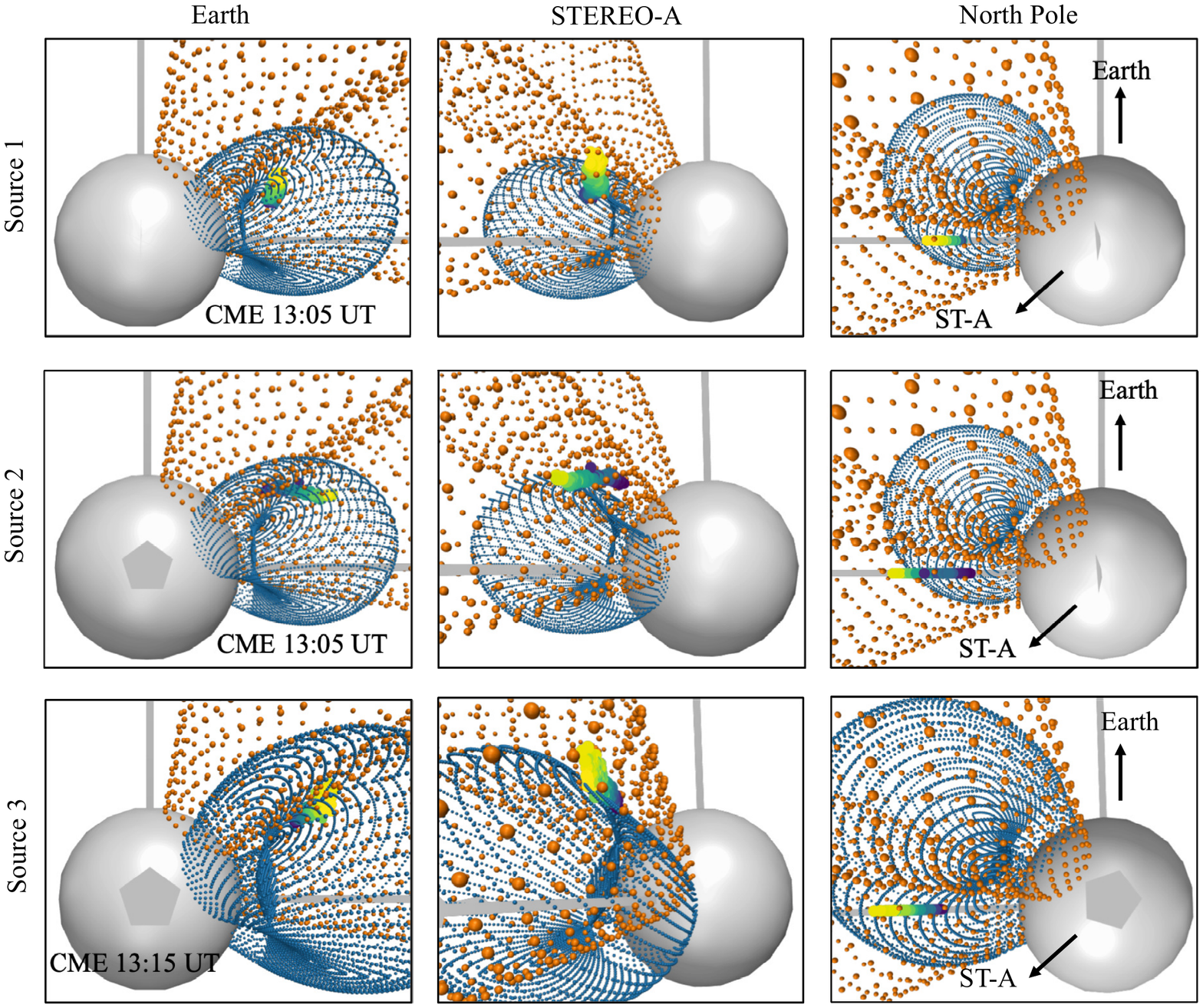}
    \caption{The CME expansion in 3D together with the associated radio emission locations. The main CME is shown as a wireframe consisting of blue dots, while the ealier CME is shown as an orange wireframe. Three viewpoints are shown, i.e.\ (left) Earth's perspective, (middle) STEREO-A's perspective, and (right) the solar North Pole.}
    \label{fig:fig8}
\end{figure*}

\begin{figure*}[ht]
\centering
    \includegraphics[width=0.9\linewidth, angle = 0, trim = {30px 0px 0px 0px}]{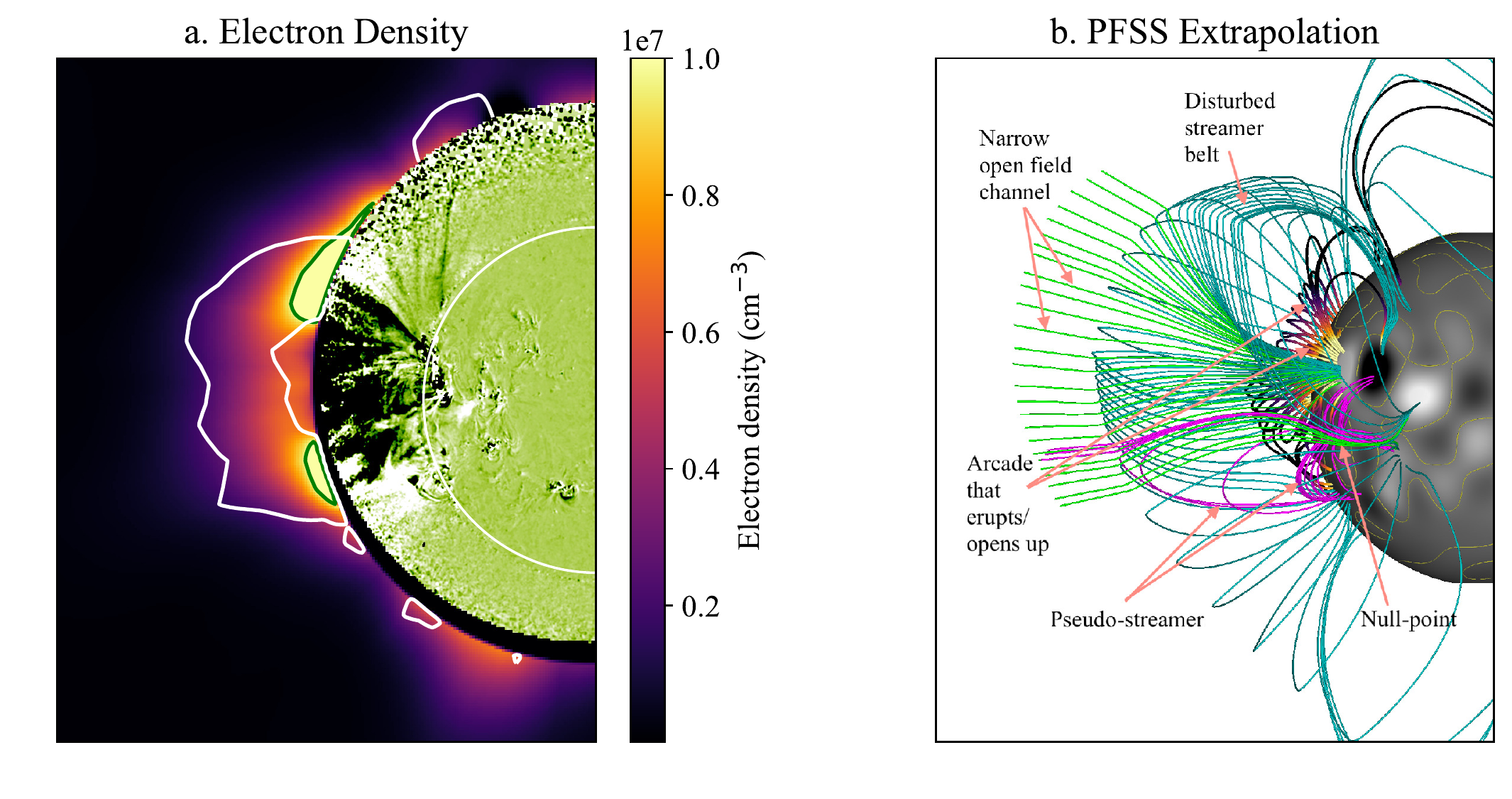}
    \caption{The electron density and magnetic field environment through which the CME propagates. (a) Electron density map of the solar corona from the STEREO-A perspective. The green contours represent the 10$^7$~cm$^{-3}$ density level and the white contours outline the CME density structure after a background map has been subtracted. 
    (b) PFSS extrapolation from the photospheric magnetic field showing the coronal flux systems in the vicinity of the CME source region.}
    \label{fig:fig9}
\end{figure*}

{The reconstructed CMEs are shown in Fig.~\ref{fig:fig7} from the Earth (left) and STEREO-A (right) perspectives, as wireframes consisting of blue dots for the main CME and orange dots for the earlier CME. The reconstructed parameters of the two CMEs, such as propagation longitude, latitude, tilt (i.e. axis orientation), and speed are given in Table~\ref{tab:tab1}. Although the two CMEs originate from the same source region (AR~11745), they have different propagation directions. The main CME expands into the earlier CME mainly with its northern portion, while the southern flank expands through the less disturbed background corona. The 3D reconstruction and coronagraph images show that the main part of the earlier CME \citep[i.e.\ the classic `three-part structure' comprising of a bright outer rim, a dark cavity corresponding to the flux rope, and a bright core;][]{il85} is already at a significant altitude when the main CME is still close to the Sun (blue wireframe). Therefore, the main CME most likely propagates partially into the legs of the previous CME. The magnetic field configuration of CME legs are usually approximately radial, however they have an additional twist component throughout their volume. While such twist is a common feature of CME models and simulations \citep[e.g.][]{li03,is07,au12,ly16,to18} there is only indirect evidence of it in solar observations \citep[e.g.][]{wa18} and in-situ observations may be difficult to interpret \citep[e.g.][]{ow16,al19}. The legs can therefore be considered as `open field lines' from the perspective of the main CME. The medium through which the CME propagates also appears significantly more disturbed at its northern flank than at its southern flank (see Figs.~\ref{fig:fig3} and \ref{fig:fig5}), which agrees with our 3D reconstruction of the geometry of the two CMEs. Bent coronal structures that may correspond to the legs of the earlier CME are indeed visible at a slightly later time in coronagraph images (see e.g.\ the features at the northern flank in Fig.~\ref{fig:fig3}a).}

{The CME reconstructions can be used to visualise the location of the associated accelerated electrons in 3D, using the \textit{x}, \textit{y}, and \textit{z}-coordinates of the radio emission centroids obtained in the previous section. The moving radio sources are shown relative to the two CMEs in 3D in Fig.~\ref{fig:fig8}. This figure represents a zoomed-in version of the perspectives in Fig.~\ref{fig:fig7}. The three moving radio sources are shown from top to bottom, and from three viewing angles: Earth (left panels), STEREO-A (middle panels), and a view from the solar North Pole (right panels). The moving radio source centroids are assumed to occur in or close to the plane of sky as outlined in the previous section. Nevertheless, we note that these radio bursts could also move away or towards the observer from Earth's viewpoint, which would translate to a movement towards Earth or a movement towards STEREO-A, respectively, in the North Pole viewpoint shown in Fig.~\ref{fig:fig8}. Studies of the Type II bursts associated with the main CME have found that the Type II emission is directed towards STEREO-A (see Fig. 2 from \citealt{ma16}). It is possible that the moving bursts studied here are also likely to be directed towards STEREO-A, away from the plane of sky, in the direction of the CME shock front expansion. The moving radio bursts positions, however, are a good representative as to where the radio bursts are located with respect to the two CMEs.  }

{All three moving radio sources appear to be located outside the reconstructed main CME bubble (blue wireframe) and at the same time inside the reconstructed CME bubble of the earlier CME (orange wireframe). Although the three moving sources are located outside the CME structure, they could still be associated to a shock or waves propagating ahead of the CME, which were not included in the wireframe reconstruction. Since the \textit{z}-coordinate of the centroids is most likely located within a fixed range close to the plane of sky, the centroid locations at the CME flank and inside the earlier CME remain unaffected by the uncertainty in the \textit{z}-coordinate. The times of the CME reconstructions labelled in Fig.~\ref{fig:fig8} are approximately during the mid-time of the propagation of the radio source centroids (which have the same colouring through time as the colour bar in Fig.~\ref{fig:fig5}) for Sources 2 and 3. In the case of Source 1, the CME reconstruction time is $\sim$18~minutes after the onset of the radio emission, since we were unable to perform a reconstruction before 13:05~UT due to the CME not fully appearing in STEREO/COR1-A (the first appearance at 12:55~UT reported in Sect.~\ref{sec:analysis} indicates the very leading edge of the CME, which is not enough to perform a full 3D reconstruction). Nevertheless, the 3D reconstructions show that all moving bursts are associated with the northern CME flank and they are located outside the main CME flank and inside the earlier CME. The 3D reconstruction also agrees with the kinematics in Fig.~\ref{fig:fig6}, where Source 1 and 3 propagate laterally outwards with the northern CME flank, while Source 2 does not show a northern propagation direction but only a radially outwards one, close to the propagation direction of the CME apex. }


\section{Discussion} \label{sec:discussion}

\begin{figure}[ht]
\centering
    \includegraphics[width=0.85\linewidth]{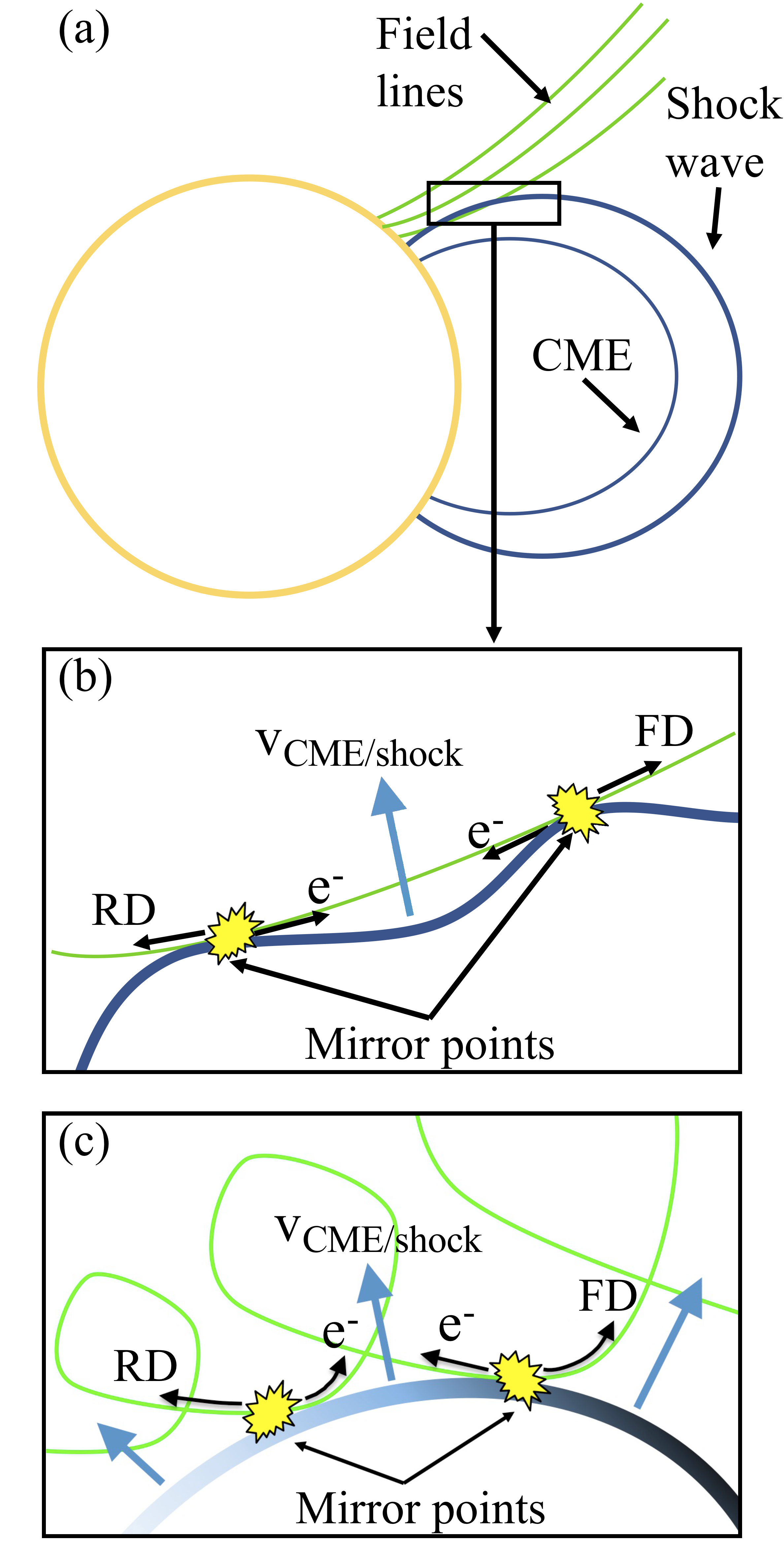}
    \caption{Schematic demonstrating the generation of the magnetic trap geometry. This schematic shows how the CME--CME or CME--background field interaction can generate collapsing magnetic trap geometries likely to generate the moving bursts studied here \citep[after][]{ma02}. When the CME-driven shock wave intersects an open field line twice, these intersections form two mirror points as the CME front advances in the direction indicated by v$_\mathrm{CME/shock}$. At the mirror points, electrons (e$^{-}$) can bounce back and forth until they escape. Once the electrons escape they can produce forward drift (FD) and reverse drift (RD) radio bursts. This can arise via a wavy shock front intersecting radial open fields (panel b) or a planar shock front intersecting helical open field lines (panel c).}
    \label{fig:fig10}
\end{figure}

{We have so far shown that bursty signatures of electron beams originate at the northern CME flank, which expands in the wake of an earlier CME. This radio emission closely follows the CME propagation. We believe that two successive CMEs created favourable conditions for the acceleration of electron beams that generated the radio emission observed at such large heights.}

{To better understand the origin of energetic electrons producing this high-altitude bursty emission, we explore in more detail the coronal densities and magnetic field environment through which the CME propagates with its northern flank. Plane-of-sky electron densities in the corona can be obtained from measurements of polarised brightness of the corona recorded by coronagraphs \citep{hu50}. The CME was first observed by STEREO/COR1-A, which can be used to construct electron density maps during its eruption, using the spherically symmetric polynomial inversion technique of \citet{wa14}, which is based on the assumption that the radial electron density distribution has a polynomial form. An example of such a map is shown in Fig.~\ref{fig:fig9}a, where the electron densities in cm$^{-3}$ obtained from COR1-A are shown in the outer field of view together with an EUVI-A image of the CME at 13:00~UT in the inner field of view for comparison. The white contours denote the outline of the CME in the density map after a background map is subtracted. The green contours denote the 10$^7$~cm$^{-3}$ density level. The moving radio bursts start propagating before the interface between the EUVI and COR1 fields of view and, in the case of Source 2 and 3, the sources propagate into the COR1 field of view before they disappear, assuming this emission occurs close to the plane of sky. A small region of enhanced densities (>10$^7$~cm$^{-3}$) at the northern CME flank in Fig.~\ref{fig:fig9}a coincides with the location of the high-altitude radio emission which also maps a high density structure. Densities >10$^7$~cm$^{-3}$ are significantly larger than those predicted by density models of the background corona such as the \citet{sa77} model or the \citet{new61} model, which would place these densities at a height of up to 1.35\,$R_\odot$. The electron density map places these densities as high as 1.7\,$R_\odot$. Such a big active region is also likely to be accompanied by enhanced densities radially outwards. We note that such density maps, as presented in Fig.~\ref{fig:fig9}a, are useful in providing a global plane-of-sky picture of the coronal electron densities. Densities of the order of 10$^7$~cm$^{-3}$ and the large heights of the radio emission observed correspond to harmonic plasma emission, since harmonic emission at a frequency of 150~MHz would originate at a plasma electron number density equal to 7$\times$10$^7$~cm$^{-3}$.}

{Models of the coronal magnetic field can be used to better understand the CME eruption and propagation through the ambient corona and its magnetic field. A selection of representative magnetic field lines associated with the flux systems in the vicinity of the CME eruption is shown in Fig.~\ref{fig:fig9}b, from the STEREO-A perspective. The Fig.~\ref{fig:fig9}b field lines are derived from the Potential Field Source Surface \citep[PFSS; e.g.][]{wa92, sc03} extrapolation based on photospheric magnetograms observed by the National Solar Observatory (NSO) Global Oscillation Network Group \citep[GONG;][]{ha96}. These field lines are a good representation of the ambient background magnetic field prior to the main CME. Close to the northern flank region, in white-light observations, we also observed the legs of the preceding CME, which can be categorised as open field with an additional twist component that would be present in addition to the PFSS field lines. }

{The field lines coloured from purple to yellow in Fig.~\ref{fig:fig9}b show the arcade above the polarity inversion line of the active region that either erupts to form the CME or opens up following the CME eruption. After eruption, the main CME propagates through a disturbed helmet streamer belt (blue field lines) at its northern flank, which changes from a north--south direction above the active region to an east--west direction north of the active region. A narrow channel of open flux is denoted by the green lines. At the southern flank, a coronal null point and pseudostreamer are present, denoted by the magenta field lines. Coronal loops that are arranged consistently with the null-point and pseudostreamer geometry can also be observed in SWAP 174~{\AA} images of the Sun before the CME eruption (see Movie~2 accompanying this paper). The radio emission originates at the northern flank, where the main CME appears to laterally expand into the closed flux system of the disturbed helmet streamer belt and the `open' field lines of the legs of the preceding CME. The pseudostreamer null point, and thus the region where magnetic reconnection would occur following the eruption, is located at the southern flank instead. Well after our three northern flank sources occur, other stationary bursts seem to indeed originate from the southern flank pseudostreamer null point in the wake of the main CME eruption, as seen in plane-of-sky images (see Movie~1 accompanying this paper.) }

{The moving radio bursts propagate outwards with the northern flank of the main CME through the magnetic field lines, strongly disturbed plasma, and possible enhanced density structures left behind by the earlier CME. An early interaction between these two CMEs is most likely the cause of this emission that shows fine structures with both forward and reverse drifts indicating individual electron beams propagating in various directions. Similar signatures of electron beams have previously been observed as herringbones, when the CME drives a shock predominantly at the flank \citep{ca87,mo19a}. However, in this case Type II emission occurs at lower frequencies during the onset of Sources 2 and 3. Thus, a shock is present in the corona at least during the occurrence of Sources 2 and 3 but it may not be the only element responsible for this emission.}

{Collapsing magnetic traps have been proposed by \citet{ma02} as a mechanism to accelerate electrons to explain the occurrence of Type II bursts and herringbones in the low corona, where shocks originate during the early stages of an eruption with low Mach numbers \citep{ku19}, making them inefficient particle accelerators. Fig.~\ref{fig:fig10}a is a schematic for possible interactions of the main CME with the magnetic field and plasma of the disturbed corona left behind by the previous CME. These interactions, illustrated in Fig.~\ref{fig:fig10}b and \ref{fig:fig10}c, are favourable conditions for the creation of collapsing magnetic traps. In this scenario, electrons can reflect at mirror points created by a shock wave intersecting magnetic field lines twice. The electrons can thus reflect several times before they become energised and escape in directions towards or away from the Sun, to generate the radio emission observed. One way for the CME-driven shock wave to intersect the open field lines twice is to have a wavy or turbulent shock front, as shown in Fig.~\ref{fig:fig10}b. An alternative, shown in Fig.\ref{fig:fig10}c, has curved or helical open magnetic field lines upstream of a planar shock \citep[e.g.][]{sa06}. The twist in the magnetic field in the legs of the earlier CME could create such conditions.}

{Theoretical studies predict that particles accelerated in magnetic traps will gain more parallel momentum, eventually being able to escape the trap, however with a small energy gain \citep{sa06}. The small energy gain suits our observations since the moving radio bursts observed consist of weak and short-lived radio emission compared to the other main radio bursts present (Type IIs and Type IVs). These temporary collapsing traps are also expected to move outwards with the CME expansion, therefore the radio bursts would show an outwards movement through time. The presence of a shock wave in the low corona at the time of the moving radio bursts is already indicated by the ongoing Type II emission lanes observed at lower frequencies. We also note that due to the presence of a disturbed streamer that has an east--west orientation at the northern flank (Fig.~\ref{fig:fig9}b), the CME propagates quasi-perpendicularly to the overlying magnetic field, making it more likely for particles to be accelerated at a CME shock front at this location. However, the presence of a CME-driven shock is not a necessary condition, since any CME-driven wave can intersect the ambient field lines in a similar manner. Such a mechanism can explain the presence of these unusual high-altitude radio bursts predominantly at the northern flank, their frequency drift, and their limited duration, since these collapsing traps disappear as the CME evolves and propagates away from the Sun.}


\section{Conclusion} \label{sec:conclusion}

{We observed for the first time bursty emission originating close to the CME flank region, which is not included in the standard classification of solar radio bursts. A likely mechanism that we propose as responsible for generating the bursty electron beams are collapsing magnetic traps. These traps are favoured by the complex ambient coronal conditions created by two interacting CMEs. Our observations show that the conditions and mechanism of particle acceleration during the CME eruption are highly dependent on the ambient plasma characteristics and complexity of the magnetic field through which the CME propagates. The collapsing magnetic trap mechanism may also be important for space weather, since it allows the acceleration of fast electrons even in weak CME-driven shocks. Collapsing magnetic traps are also a universal acceleration mechanism, since they can be found in other plasmas such as planetary magnetospheres and bow shocks \citep[e.g.][]{gi90}.}

{Our study demonstrates that the additional perspectives provided by the STEREO mission were necessary to determine the nature of the moving radio bursts. 3D reconstructions of the eruptions were essential to show how the two CMEs propagate with respect to each other. Future observations of moving radio bursts associated with CMEs, combined with possibly additional vantage points from the STEREO-A spacecraft and future L5 missions, should be used to better understand the numerous possibilities for particle acceleration during CME eruptions. Such studies also have the potential to update the current classification of solar radio emission.}


\begin{acknowledgements}{The results presented here have been achieved under the framework of the Finnish Centre of Excellence in Research of Sustainable Space (Academy of Finland grant number 312390), which we gratefully acknowledge. E.P. acknowledges the NASA Living With a Star Jack Eddy Postdoctoral Fellowship Program, administered by UCAR's Cooperative Programs for the Advancement of Earth System Science (CPAESS) under award no. NNX16AK22G. E.K.J.K. acknowledges the ERC under the European Union's Horizon 2020 Research and Innovation Programme Project SolMAG 724391, and Academy of Finland Project 310445. J.M. acknowledges funding by the  BRAIN-be (Belgian Research Action through Interdisciplinary Networks) project CCSOM (Constraining CMEs and Shocks by Observations and Modelling throughout the inner heliosphere). B.J.L. acknowledges support from NASA HGI 80NSSC18K0645, LWS 80NSSC19K0088, and NSF AGS 1851945. M.P. acknowledges the European Research Council (ERC) Consolidator grant 682068-PRESTISSIMO, and Academy of Finland grants 312351 and 309937. We thank the Radio Solar Database service at LESIA \& USN (Observatoire de Paris) for making the NRH/NDA/ORFEES data available. }\end{acknowledgements}

\bibliographystyle{aa} 
\bibliography{aanda.bib} 

\begin{thebibliography}{82}
\expandafter\ifx\csname natexlab\endcsname\relax\def\natexlab#1{#1}\fi

\bibitem[{{Al-Haddad} {et~al.}(2019){Al-Haddad}, {Poedts}, {Roussev},
  {Farrugia}, {Yu}, \& {Lugaz}}]{al19}
{Al-Haddad}, N., {Poedts}, S., {Roussev}, I., {et~al.} 2019, \apj, 870, 100

\bibitem[{{Aulanier} {et~al.}(2012){Aulanier}, {Janvier}, \&
  {Schmieder}}]{au12}
{Aulanier}, G., {Janvier}, M., \& {Schmieder}, B. 2012, \aap, 543, A110

\bibitem[{{Bastian} {et~al.}(2001){Bastian}, {Pick}, {Kerdraon}, {Maia}, \&
  {Vourlidas}}]{ba01}
{Bastian}, T.~S., {Pick}, M., {Kerdraon}, A., {Maia}, D., \& {Vourlidas}, A.
  2001, \apjl, 558, L65

\bibitem[{{Boischot}(1957)}]{bo57}
{Boischot}, A. 1957, Academie des Sciences Paris Comptes Rendus, 244, 1326

\bibitem[{{Boischot} {et~al.}(1980){Boischot}, {Rosolen}, {Aubier}, {Daigne},
  {Genova}, {Leblanc}, {Lecacheux}, {de La Noe}, \& {Moller-Pedersen}}]{bo80}
{Boischot}, A., {Rosolen}, C., {Aubier}, M.~G., {et~al.} 1980, \icarus, 43, 399

\bibitem[{{Bougeret} {et~al.}(2008){Bougeret}, {Goetz}, {Kaiser}, {Bale},
  {Kellogg}, {Maksimovic}, {Monge}, {Monson}, {Astier}, {Davy}, {Dekkali},
  {Hinze}, {Manning}, {Aguilar-Rodriguez}, {Bonnin}, {Briand}, {Cairns},
  {Cattell}, {Cecconi}, {Eastwood}, {Ergun}, {Fainberg}, {Hoang}, {Huttunen},
  {Krucker}, {Lecacheux}, {MacDowall}, {Macher}, {Mangeney}, {Meetre},
  {Moussas}, {Nguyen}, {Oswald}, {Pulupa}, {Reiner}, {Robinson}, {Rucker},
  {Salem}, {Santolik}, {Silvis}, {Ullrich}, {Zarka}, \&
  {Zouganelis}}]{bougeret2008}
{Bougeret}, J.~L., {Goetz}, K., {Kaiser}, M.~L., {et~al.} 2008, \ssr, 136, 487

\bibitem[{{Bouratzis} {et~al.}(2016){Bouratzis}, {Hillaris}, {Alissandrakis},
  {Preka-Papadema}, {Moussas}, {Caroubalos}, {Tsitsipis}, \&
  {Kontogeorgos}}]{bo16}
{Bouratzis}, C., {Hillaris}, A., {Alissandrakis}, C.~E., {et~al.} 2016, \aap,
  586, A29

\bibitem[{{Brueckner} {et~al.}(1995){Brueckner}, {Howard}, {Koomen},
  {Korendyke}, {Michels}, {Moses}, {Socker}, {Dere}, {Lamy}, {Llebaria},
  {Bout}, {Schwenn}, {Simnett}, {Bedford}, \& {Eyles}}]{br95}
{Brueckner}, G.~E., {Howard}, R.~A., {Koomen}, M.~J., {et~al.} 1995, \solphys,
  162, 357

\bibitem[{{Cairns} \& {Robinson}(1987)}]{ca87}
{Cairns}, I.~H. \& {Robinson}, R.~D. 1987, \solphys, 111, 365

\bibitem[{{Cane} \& {White}(1989)}]{ca89}
{Cane}, H.~V. \& {White}, S.~M. 1989, \solphys, 120, 137

\bibitem[{{Carley} {et~al.}(2013){Carley}, {Long}, {Byrne}, {Zucca},
  {Bloomfield}, {McCauley}, \& {Gallagher}}]{ca13}
{Carley}, E.~P., {Long}, D.~M., {Byrne}, J.~P., {et~al.} 2013, Nat. Phys., 9,
  811

\bibitem[{{Chrysaphi} {et~al.}(2018){Chrysaphi}, {Kontar}, {Holman}, \&
  {Temmer}}]{chr18}
{Chrysaphi}, N., {Kontar}, E.~P., {Holman}, G.~D., \& {Temmer}, M. 2018, \apj,
  868, 79

\bibitem[{{Chrysaphi} {et~al.}(2020){Chrysaphi}, {Reid}, \& {Kontar}}]{chr20}
{Chrysaphi}, N., {Reid}, H. A.~S., \& {Kontar}, E.~P. 2020, \apj, 893, 115

\bibitem[{Ding {et~al.}(2013)Ding, Jiang, Zhao, \& Li}]{di13}
Ding, L., Jiang, Y., Zhao, L., \& Li, G. 2013, The Astrophysical Journal, 763,
  30

\bibitem[{{Domingo} {et~al.}(1995){Domingo}, {Fleck}, \& {Poland}}]{do95}
{Domingo}, V., {Fleck}, B., \& {Poland}, A.~I. 1995, \solphys, 162, 1

\bibitem[{{Dulk}(1973)}]{du73}
{Dulk}, G.~A. 1973, \solphys, 32, 491

\bibitem[{{Gisler} \& {Lemons}(1990)}]{gi90}
{Gisler}, G. \& {Lemons}, D. 1990, \jgr, 95, 14925

\bibitem[{{Gopalswamy} {et~al.}(2016){Gopalswamy}, {Yashiro}, {Thakur},
  {M{\"a}kel{\"a}}, {Xie}, \& {Akiyama}}]{go16}
{Gopalswamy}, N., {Yashiro}, S., {Thakur}, N., {et~al.} 2016, \apj, 833, 216

\bibitem[{{Harvey} {et~al.}(1996){Harvey}, {Hill}, {Hubbard}, {Kennedy},
  {Leibacher}, {Pintar}, {Gilman}, {Noyes}, {Title}, {Toomre}, {Ulrich},
  {Bhatnagar}, {Kennewell}, {Marquette}, {Patron}, {Saa}, \& {Yasukawa}}]{ha96}
{Harvey}, J.~W., {Hill}, F., {Hubbard}, R.~P., {et~al.} 1996, Science, 272,
  1284

\bibitem[{{Holman} \& {Pesses}(1983)}]{ho83}
{Holman}, G.~D. \& {Pesses}, M.~E. 1983, \apj, 267, 837

\bibitem[{{Howard} {et~al.}(2008){Howard}, {Moses}, {Vourlidas}, {Newmark},
  {Socker}, {Plunkett}, {Korendyke}, {Cook}, {Hurley}, {Davila}, {Thompson},
  {St Cyr}, {Mentzell}, {Mehalick}, {Lemen}, {Wuelser}, {Duncan}, {Tarbell},
  {Wolfson}, {Moore}, {Harrison}, {Waltham}, {Lang}, {Davis}, {Eyles},
  {Mapson-Menard}, {Simnett}, {Halain}, {Defise}, {Mazy}, {Rochus}, {Mercier},
  {Ravet}, {Delmotte}, {Auchere}, {Delaboudiniere}, {Bothmer}, {Deutsch},
  {Wang}, {Rich}, {Cooper}, {Stephens}, {Maahs}, {Baugh}, {McMullin}, \&
  {Carter}}]{ho08}
{Howard}, R.~A., {Moses}, J.~D., {Vourlidas}, A., {et~al.} 2008, \ssr, 136, 67

\bibitem[{{Illing} \& {Hundhausen}(1985)}]{il85}
{Illing}, R.~M.~E. \& {Hundhausen}, A.~J. 1985, J. Geophys. Res. Space Phys.,
  90, 275

\bibitem[{{Isenberg} \& {Forbes}(2007)}]{is07}
{Isenberg}, P.~A. \& {Forbes}, T.~G. 2007, \apj, 670, 1453

\bibitem[{{Kahler} \& {Hundhausen}(1992)}]{ka92}
{Kahler}, S.~W. \& {Hundhausen}, A.~J. 1992, \jgr, 97, 1619

\bibitem[{{Kaiser} {et~al.}(2008){Kaiser}, {Kucera}, {Davila}, {St.~Cyr},
  {Guhathakurta}, \& {Christian}}]{ka08}
{Kaiser}, M.~L., {Kucera}, T.~A., {Davila}, J.~M., {et~al.} 2008, \ssr, 136, 5

\bibitem[{{Kerdraon} \& {Delouis}(1997)}]{ke97}
{Kerdraon}, A. \& {Delouis}, J.-M. 1997, in Lecture Notes in Physics, Berlin
  Springer Verlag, Vol. 483, Coronal Physics from Radio and Space Observations,
  ed. G.~{Trottet}, 192

\bibitem[{{Klassen} {et~al.}(2002){Klassen}, {Bothmer}, {Mann}, {Reiner},
  {Krucker}, {Vourlidas}, \& {Kunow}}]{kl02}
{Klassen}, A., {Bothmer}, V., {Mann}, G., {et~al.} 2002, \aap, 385, 1078

\bibitem[{{Kong} {et~al.}(2012){Kong}, {Chen}, {Li}, {Feng}, {Song}, {Guo}, \&
  {Jiao}}]{ko12}
{Kong}, X.~L., {Chen}, Y., {Li}, G., {et~al.} 2012, \apj, 750, 158

\bibitem[{Kumari {et~al.}(2019)Kumari, Ramesh, Kathiravan, Wang, \&
  Gopalswamy}]{ku19}
Kumari, A., Ramesh, R., Kathiravan, C., Wang, T.~J., \& Gopalswamy, N. 2019,
  The Astrophysical Journal, 881, 24

\bibitem[{{Kwon} {et~al.}(2014){Kwon}, {Zhang}, \& {Olmedo}}]{kw14}
{Kwon}, R.-Y., {Zhang}, J., \& {Olmedo}, O. 2014, \apj, 794, 148

\bibitem[{{Lemen} {et~al.}(2012){Lemen}, {Title}, {Akin}, {Boerner}, {Chou},
  {Drake}, {Duncan}, {Edwards}, {Friedlaender}, {Heyman}, {Hurlburt}, {Katz},
  {Kushner}, {Levay}, {Lindgren}, {Mathur}, {McFeaters}, {Mitchell}, {Rehse},
  {Schrijver}, {Springer}, {Stern}, {Tarbell}, {Wuelser}, {Wolfson}, {Yanari},
  {Bookbinder}, {Cheimets}, {Caldwell}, {Deluca}, {Gates}, {Golub}, {Park},
  {Podgorski}, {Bush}, {Scherrer}, {Gummin}, {Smith}, {Auker}, {Jerram},
  {Pool}, {Soufli}, {Windt}, {Beardsley}, {Clapp}, {Lang}, \& {Waltham}}]{le12}
{Lemen}, J.~R., {Title}, A.~M., {Akin}, D.~J., {et~al.} 2012, \solphys, 275, 17

\bibitem[{{Linker} {et~al.}(2003){Linker}, {Miki{\'c}}, {Lionello}, {Riley},
  {Amari}, \& {Odstrcil}}]{li03}
{Linker}, J.~A., {Miki{\'c}}, Z., {Lionello}, R., {et~al.} 2003, Phys. Plasmas,
  10, 1971

\bibitem[{{Liu} {et~al.}(2009){Liu}, {Luhmann}, {Bale}, \& {Lin}}]{liu09}
{Liu}, Y., {Luhmann}, J.~G., {Bale}, S.~D., \& {Lin}, R.~P. 2009, \apjl, 691,
  L151

\bibitem[{{Liu} {et~al.}(2017){Liu}, {Hu}, {Zhu}, {Luhmann}, \&
  {Vourlidas}}]{liu17}
{Liu}, Y.~D., {Hu}, H., {Zhu}, B., {Luhmann}, J.~G., \& {Vourlidas}, A. 2017,
  \apj, 834, 158

\bibitem[{{Liu} {et~al.}(2013){Liu}, {Luhmann}, {Lugaz}, {M{\"o}stl}, {Davies},
  {Bale}, \& {Lin}}]{liu13}
{Liu}, Y.~D., {Luhmann}, J.~G., {Lugaz}, N., {et~al.} 2013, \apj, 769, 45

\bibitem[{{Liu} {et~al.}(2019){Liu}, {Zhu}, \& {Zhao}}]{liu19}
{Liu}, Y.~D., {Zhu}, B., \& {Zhao}, X. 2019, \apj, 871, 8

\bibitem[{{Lynch} {et~al.}(2016){Lynch}, {Masson}, {Li}, {DeVore}, {Luhmann},
  {Antiochos}, \& {Fisher}}]{ly16}
{Lynch}, B.~J., {Masson}, S., {Li}, Y., {et~al.} 2016, \jgr, 121, 10,677

\bibitem[{{Magdaleni{\'c}} {et~al.}(2020){Magdaleni{\'c}}, {Marqu{\'e}},
  {Fallows}, {Mann}, {Vocks}, {Zucca}, {Dabrowski}, {Krankowski}, \&
  {Melnik}}]{ma20}
{Magdaleni{\'c}}, J., {Marqu{\'e}}, C., {Fallows}, R.~A., {et~al.} 2020, \apjl,
  897, L15

\bibitem[{{Magdaleni{\'c}} {et~al.}(2014){Magdaleni{\'c}}, {Marqu{\'e}},
  {Krupar}, {Mierla}, {Zhukov}, {Rodriguez}, {Maksimovi{\'c}}, \&
  {Cecconi}}]{ma14}
{Magdaleni{\'c}}, J., {Marqu{\'e}}, C., {Krupar}, V., {et~al.} 2014, \apj, 791,
  115

\bibitem[{{Magdaleni{\'c}} {et~al.}(2002){Magdaleni{\'c}}, {Vr{\v{s}}nak}, \&
  {Aurass}}]{ma02}
{Magdaleni{\'c}}, J., {Vr{\v{s}}nak}, B., \& {Aurass}, H. 2002, in ESA Special
  Publication, Vol.~1, Solar Variability: From Core to Outer Frontiers, ed.
  A.~{Wilson}, 335--338

\bibitem[{{Magdaleni{\'c}} {et~al.}(2006){Magdaleni{\'c}}, {Vr{\v{s}}nak},
  {Zlobec}, {Hillaris}, \& {Messerotti}}]{ma06}
{Magdaleni{\'c}}, J., {Vr{\v{s}}nak}, B., {Zlobec}, P., {Hillaris}, A., \&
  {Messerotti}, M. 2006, \apjl, 642, L77

\bibitem[{{Maguire} {et~al.}(2020){Maguire}, {Carley}, {McCauley}, \&
  {Gallagher}}]{maguire20}
{Maguire}, C.~A., {Carley}, E.~P., {McCauley}, J., \& {Gallagher}, P.~T. 2020,
  \aap, 633, A56

\bibitem[{{M{\"a}kel{\"a}} {et~al.}(2016){M{\"a}kel{\"a}}, {Gopalswamy},
  {Reiner}, {Akiyama}, \& {Krupar}}]{ma16}
{M{\"a}kel{\"a}}, P., {Gopalswamy}, N., {Reiner}, M.~J., {Akiyama}, S., \&
  {Krupar}, V. 2016, \apj, 827, 141

\bibitem[{{Mancuso} {et~al.}(2019){Mancuso}, {Frassati}, {Bemporad}, \&
  {Barghini}}]{ma19}
{Mancuso}, S., {Frassati}, F., {Bemporad}, A., \& {Barghini}, D. 2019, \aap,
  624, L2

\bibitem[{{Mancuso} \& {Raymond}(2004)}]{ma04}
{Mancuso}, S. \& {Raymond}, J.~C. 2004, \aap, 413, 363

\bibitem[{{Mann} {et~al.}(1996){Mann}, {Klassen}, {Classen}, {Aurass},
  {Scholz}, {MacDowall}, \& {Stone}}]{ma96}
{Mann}, G., {Klassen}, A., {Classen}, H.~T., {et~al.} 1996, \aaps, 119, 489

\bibitem[{{Mart{\'\i}nez Oliveros} {et~al.}(2012){Mart{\'\i}nez Oliveros},
  {Raftery}, {Bain}, {Liu}, {Krupar}, {Bale}, \& {Krucker}}]{ma12}
{Mart{\'\i}nez Oliveros}, J.~C., {Raftery}, C.~L., {Bain}, H.~M., {et~al.}
  2012, \apj, 748, 66

\bibitem[{{Morosan} {et~al.}(2019{\natexlab{a}}){Morosan}, {Carley}, {Hayes},
  {Murray}, {Zucca}, {Fallows}, {McCauley}, {Kilpua}, {Mann}, {Vocks}, \&
  {Gallagher}}]{mo19a}
{Morosan}, D.~E., {Carley}, E.~P., {Hayes}, L.~A., {et~al.} 2019{\natexlab{a}},
  Nat. Astron., 3, 452

\bibitem[{{Morosan} {et~al.}(2019{\natexlab{b}}){Morosan}, {Kilpua}, {Carley},
  \& {Monstein}}]{mo19b}
{Morosan}, D.~E., {Kilpua}, E.~K.~J., {Carley}, E.~P., \& {Monstein}, C.
  2019{\natexlab{b}}, \aap, 623, A63

\bibitem[{{Morosan} {et~al.}(2020{\natexlab{a}}){Morosan}, {Palmerio}, {Lynch},
  \& {Kilpua}}]{mo20a}
{Morosan}, D.~E., {Palmerio}, E., {Lynch}, B.~J., \& {Kilpua}, E.~K.~J.
  2020{\natexlab{a}}, \aap, 633, A141

\bibitem[{{Morosan} {et~al.}(2020{\natexlab{b}}){Morosan}, {Palmerio},
  {Pomoell}, {Vainio}, {Palmroth}, \& {Kilpua}}]{mo20b}
{Morosan}, D.~E., {Palmerio}, E., {Pomoell}, J., {et~al.} 2020{\natexlab{b}},
  \aap, 635, A62

\bibitem[{{Nelson} \& {Melrose}(1985)}]{ne85}
{Nelson}, G.~J. \& {Melrose}, D.~B. 1985, in IN: Solar radiophysics: Studies of
  emission from the sun at metre wavelengths (A87-13851 03-92). Cambridge and
  New York, Cambridge University Press, 1985, p. 333-359., ed. D.~J. {McLean}
  \& N.~R. {Labrum}, 333--359

\bibitem[{{Newkirk}(1961)}]{new61}
{Newkirk}, G.~J. 1961, \apj, 133, 983

\bibitem[{{Ogilvie} \& {Desch}(1997)}]{og97}
{Ogilvie}, K.~W. \& {Desch}, M.~D. 1997, Adv. Space Res., 20, 559

\bibitem[{{Owens}(2016)}]{ow16}
{Owens}, M.~J. 2016, \apj, 818, 197

\bibitem[{{Palmerio} {et~al.}(2019){Palmerio}, {Scolini}, {Barnes},
  {Magdaleni{\'c}}, {West}, {Zhukov}, {Rodriguez}, {Mierla}, {Good}, {Morosan},
  {Kilpua}, {Pomoell}, \& {Poedts}}]{pa19}
{Palmerio}, E., {Scolini}, C., {Barnes}, D., {et~al.} 2019, \apj, 878, 37

\bibitem[{{Pesnell} {et~al.}(2012){Pesnell}, {Thompson}, \&
  {Chamberlin}}]{pe12}
{Pesnell}, W.~D., {Thompson}, B.~J., \& {Chamberlin}, P.~C. 2012, \solphys,
  275, 3

\bibitem[{{Pomoell} {et~al.}(2008){Pomoell}, {Vainio}, \& {Kissmann}}]{po08}
{Pomoell}, J., {Vainio}, R., \& {Kissmann}, R. 2008, \solphys, 253, 249

\bibitem[{{Reiner} {et~al.}(2007){Reiner}, {Kaiser}, \& {Bougeret}}]{re07}
{Reiner}, M.~J., {Kaiser}, M.~L., \& {Bougeret}, J.~L. 2007, \apj, 663, 1369

\bibitem[{{Richardson} {et~al.}(2003){Richardson}, {Lawrence}, {Haggerty},
  {Kucera}, \& {Szabo}}]{ri03}
{Richardson}, I.~G., {Lawrence}, G.~R., {Haggerty}, D.~K., {Kucera}, T.~A., \&
  {Szabo}, A. 2003, \grl, 30, 8014

\bibitem[{{Saito} {et~al.}(1977){Saito}, {Poland}, \& {Munro}}]{sa77}
{Saito}, K., {Poland}, A.~I., \& {Munro}, R.~H. 1977, \solphys, 55, 121

\bibitem[{{Sandroos} \& {Vainio}(2006)}]{sa06}
{Sandroos}, A. \& {Vainio}, R. 2006, \aap, 455, 685

\bibitem[{{Santandrea} {et~al.}(2013){Santandrea}, {Gantois}, {Strauch},
  {Teston}, {Tilmans}, {Baijot}, {Gerrits}, {De Groof}, {Schwehm}, \&
  {Zender}}]{sa13}
{Santandrea}, S., {Gantois}, K., {Strauch}, K., {et~al.} 2013, \solphys, 286, 5

\bibitem[{{Schrijver} \& {De Rosa}(2003)}]{sc03}
{Schrijver}, C.~J. \& {De Rosa}, M.~L. 2003, \solphys, 212, 165

\bibitem[{{Seaton} {et~al.}(2013){Seaton}, {Berghmans}, {Nicula}, {Halain}, {De
  Groof}, {Thibert}, {Bloomfield}, {Raftery}, {Gallagher}, {Auch{\`e}re},
  {Defise}, {D'Huys}, {Lecat}, {Mazy}, {Rochus}, {Rossi}, {Sch{\"u}hle},
  {Slemzin}, {Yalim}, \& {Zender}}]{se13}
{Seaton}, D.~B., {Berghmans}, D., {Nicula}, B., {et~al.} 2013, \solphys, 286,
  43

\bibitem[{{Stewart} {et~al.}(1978){Stewart}, {Duncan}, {Suzuki}, \&
  {Nelson}}]{st78}
{Stewart}, R.~T., {Duncan}, R.~A., {Suzuki}, S., \& {Nelson}, G.~J. 1978, Proc.
  Astron. Soc. Australia, 3, 247

\bibitem[{{Stewart} {et~al.}(1974){Stewart}, {Howard}, {Hansen}, {Gergely}, \&
  {Kundu}}]{st74}
{Stewart}, R.~T., {Howard}, R.~A., {Hansen}, F., {Gergely}, T., \& {Kundu}, M.
  1974, \solphys, 36, 219

\bibitem[{{Thernisien} {et~al.}(2009){Thernisien}, {Vourlidas}, \&
  {Howard}}]{the09}
{Thernisien}, A., {Vourlidas}, A., \& {Howard}, R.~A. 2009, \solphys, 256, 111

\bibitem[{{Thernisien} {et~al.}(2006){Thernisien}, {Howard}, \&
  {Vourlidas}}]{the06}
{Thernisien}, A.~F.~R., {Howard}, R.~A., \& {Vourlidas}, A. 2006, \apj, 652,
  763

\bibitem[{{T{\"o}r{\"o}k} {et~al.}(2018){T{\"o}r{\"o}k}, {Downs}, {Linker},
  {Lionello}, {Titov}, {Miki{\'c}}, {Riley}, {Caplan}, \& {Wijaya}}]{to18}
{T{\"o}r{\"o}k}, T., {Downs}, C., {Linker}, J.~A., {et~al.} 2018, \apj, 856, 75

\bibitem[{{van de Hulst}(1950)}]{hu50}
{van de Hulst}, H.~C. 1950, \apj, 112, 1

\bibitem[{{Vasanth} {et~al.}(2019){Vasanth}, {Chen}, {Lv}, {Ning}, {Li},
  {Feng}, {Wu}, \& {Du}}]{va19}
{Vasanth}, V., {Chen}, Y., {Lv}, M., {et~al.} 2019, \apj, 870, 30

\bibitem[{{Vourlidas} {et~al.}(2013){Vourlidas}, {Lynch}, {Howard}, \&
  {Li}}]{vo13}
{Vourlidas}, A., {Lynch}, B.~J., {Howard}, R.~A., \& {Li}, Y. 2013, \solphys,
  284, 179

\bibitem[{{Vourlidas} {et~al.}(2003){Vourlidas}, {Wu}, {Wang}, {Subramanian},
  \& {Howard}}]{vo03}
{Vourlidas}, A., {Wu}, S.~T., {Wang}, A.~H., {Subramanian}, P., \& {Howard},
  R.~A. 2003, \apj, 598, 1392

\bibitem[{{Vr{\v s}nak} {et~al.}(2003){Vr{\v s}nak}, {Klein}, {Warmuth},
  {Otruba}, \& {Skender}}]{vr03}
{Vr{\v s}nak}, B., {Klein}, K.-L., {Warmuth}, A., {Otruba}, W., \& {Skender},
  M. 2003, \solphys, 214, 325

\bibitem[{{Vr{\v{s}}nak} {et~al.}(2001){Vr{\v{s}}nak}, {Aurass},
  {Magdaleni{\'c}}, \& {Gopalswamy}}]{vr01}
{Vr{\v{s}}nak}, B., {Aurass}, H., {Magdaleni{\'c}}, J., \& {Gopalswamy}, N.
  2001, \aap, 377, 321

\bibitem[{{Vr{\v{s}}nak} {et~al.}(2002){Vr{\v{s}}nak}, {Magdaleni{\'c}},
  {Aurass}, \& {Mann}}]{vr02}
{Vr{\v{s}}nak}, B., {Magdaleni{\'c}}, J., {Aurass}, H., \& {Mann}, G. 2002,
  \aap, 396, 673

\bibitem[{{Wang} \& {Davila}(2014)}]{wa14}
{Wang}, T. \& {Davila}, J.~M. 2014, \solphys, 289, 3723

\bibitem[{{Wang} \& {Hess}(2018)}]{wa18}
{Wang}, Y.~M. \& {Hess}, P. 2018, \apj, 853, 103

\bibitem[{{Wang} \& {Sheeley}(1992)}]{wa92}
{Wang}, Y.-M. \& {Sheeley}, N.~R., J. 1992, \apj, 392, 310

\bibitem[{{Zhao} {et~al.}(2019){Zhao}, {Liu}, {Hu}, \& {Wang}}]{zh19}
{Zhao}, X., {Liu}, Y.~D., {Hu}, H., \& {Wang}, R. 2019, \apj, 882, 122

\bibitem[{{Zucca} {et~al.}(2018){Zucca}, {Morosan}, {Rouillard}, {Fallows},
  {Gallagher}, {Magdalenic}, {Klein}, {Mann}, {Vocks}, {Carley}, {Bisi},
  {Kontar}, {Rothkaehl}, {Dabrowski}, {Krankowski}, {Anderson}, {Asgekar},
  {Bell}, {Bentum}, {Best}, {Blaauw}, {Breitling}, {Broderick}, {Brouw},
  {Br{\"u}ggen}, {Butcher}, {Ciardi}, {de Geus}, {Deller}, {Duscha},
  {Eisl{\"o}ffel}, {Garrett}, {Grie{\ss}meier}, {Gunst}, {Heald}, {Hoeft},
  {H{\"o}randel}, {Iacobelli}, {Juette}, {Karastergiou}, {van Leeuwen},
  {McKay-Bukowski}, {Mulder}, {Munk}, {Nelles}, {Orru}, {Paas}, {Pand ey},
  {Pekal}, {Pizzo}, {Polatidis}, {Reich}, {Rowlinson}, {Schwarz}, {Shulevski},
  {Sluman}, {Smirnov}, {Sobey}, {Soida}, {Thoudam}, {Toribio}, {Vermeulen},
  {van Weeren}, {Wucknitz}, \& {Zarka}}]{zu18}
{Zucca}, P., {Morosan}, D.~E., {Rouillard}, A.~P., {et~al.} 2018, \aap, 615,
  A89

\end{thebibliography}

\begin{appendix} 

\section{Spectral properties of the moving radio bursts} \label{app:a}

\begin{figure}[h]
\centering
    \includegraphics[width=0.98\linewidth]{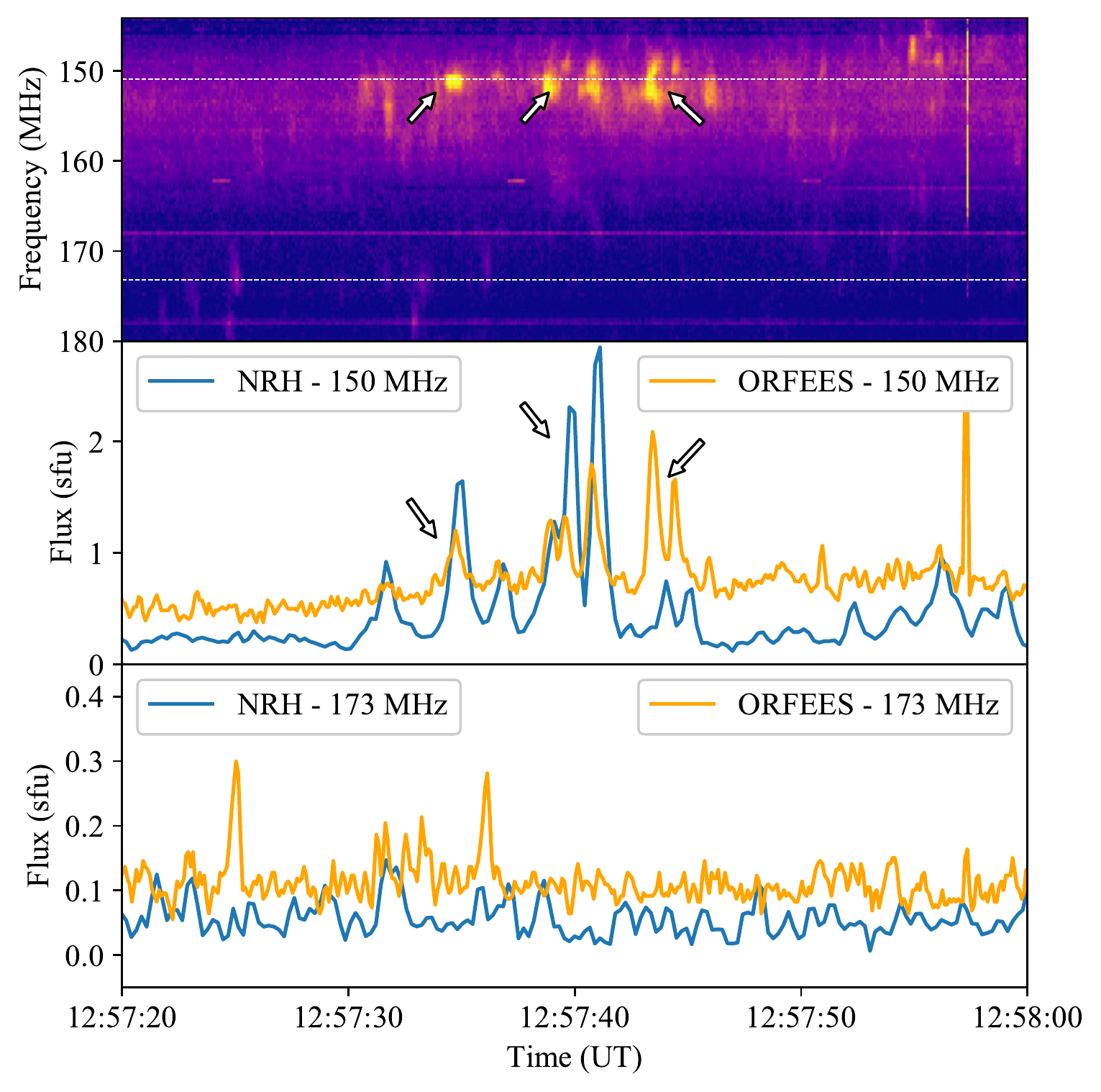}
    \caption{Spectral features corresponding to Source 1. (a) Zoomed-in view of the NDA dynamic spectrum showing a 40-s time period starting from 12:57:20~UT. (b) Time series of the NRH flux density of Source 1 in sfu and ORFEES normalised intensity in arbitrary units at 150.9~MHz. (c) Time series of the NRH flux density of Source 1 in sfu and ORFEES normalised intensity in arbitrary units at 150.9~MHz.}
    \label{fig:figa1}
\end{figure}

{The moving radio bursts presented in this study correspond to fine structures in dynamic spectra observed by ORFEES. In this appendix, we show characteristic dynamic spectra for the three moving sources where we found fine-structured bursts in ORFEES corresponding to the NRH moving bursts intensity peaks. The ORFEES dynamic spectra, together with the NRH flux densities and ORFEES time series of the normalised intensity at 150 and 173~MHz, are shown in Figs.~\ref{fig:figa1}, \ref{fig:figa1}, and \ref{fig:figa3} corresponding to Sources 1, 2, and 3, respectively. The white arrows in each figure point to the fine structures corresponding to the moving radio bursts. }

\begin{figure}
\centering
    \includegraphics[width=0.98\linewidth]{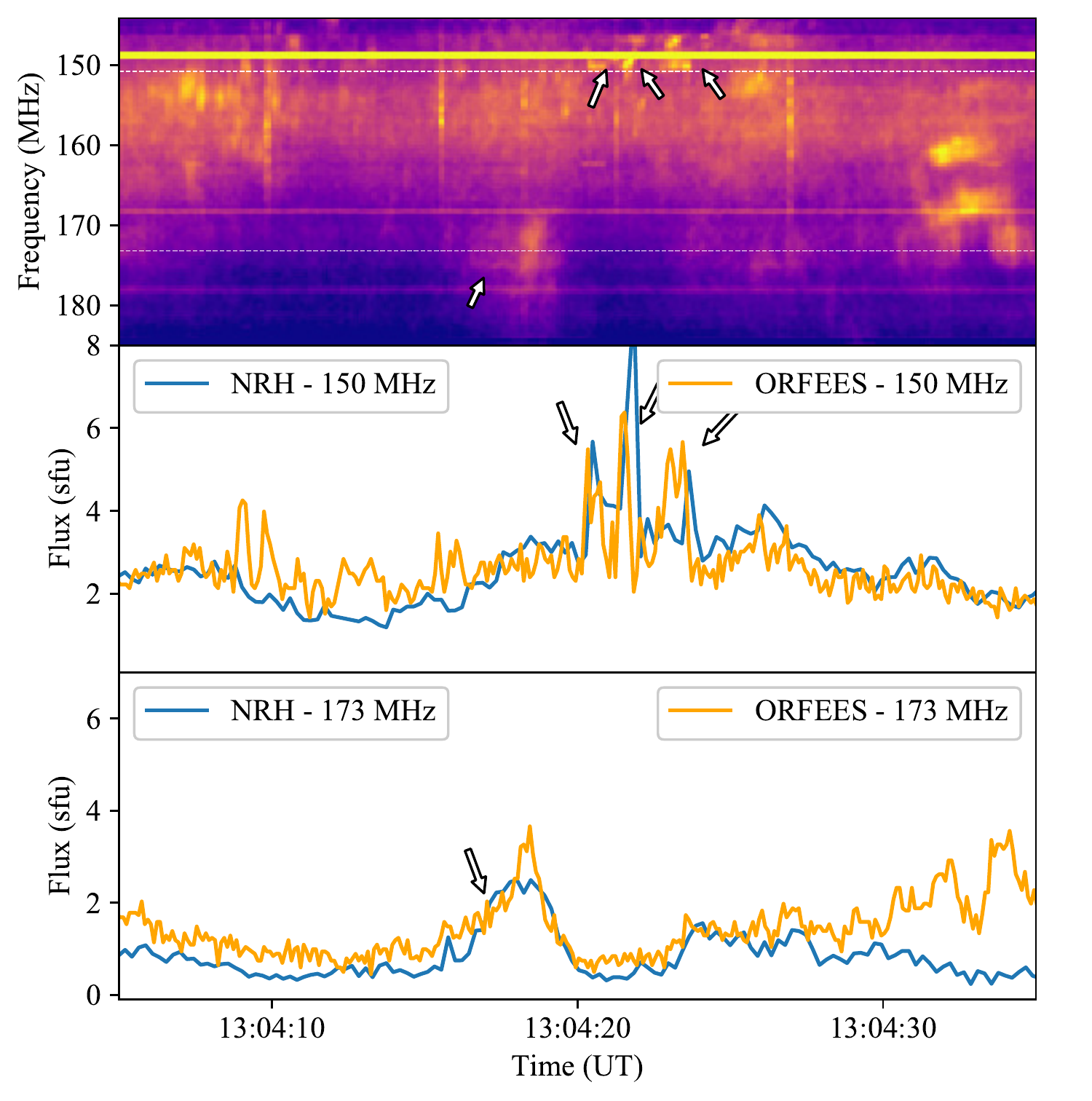}
    \caption{Spectral features corresponding to Source 2. (a) Zoomed-in view of the NDA dynamic spectrum showing a 30-s time period starting from 13:04:05~UT. (b) Time series of the NRH flux density of Source 2 in sfu and ORFEES normalised intensity in arbitrary units at 150.9~MHz. (c) Time series of the NRH flux density of Source 2 in sfu and ORFEES normalised intensity in arbitrary units at 150.9~MHz.}
    \label{fig:figa2}
\end{figure}

\begin{figure}
\centering
    \includegraphics[width=0.98\linewidth]{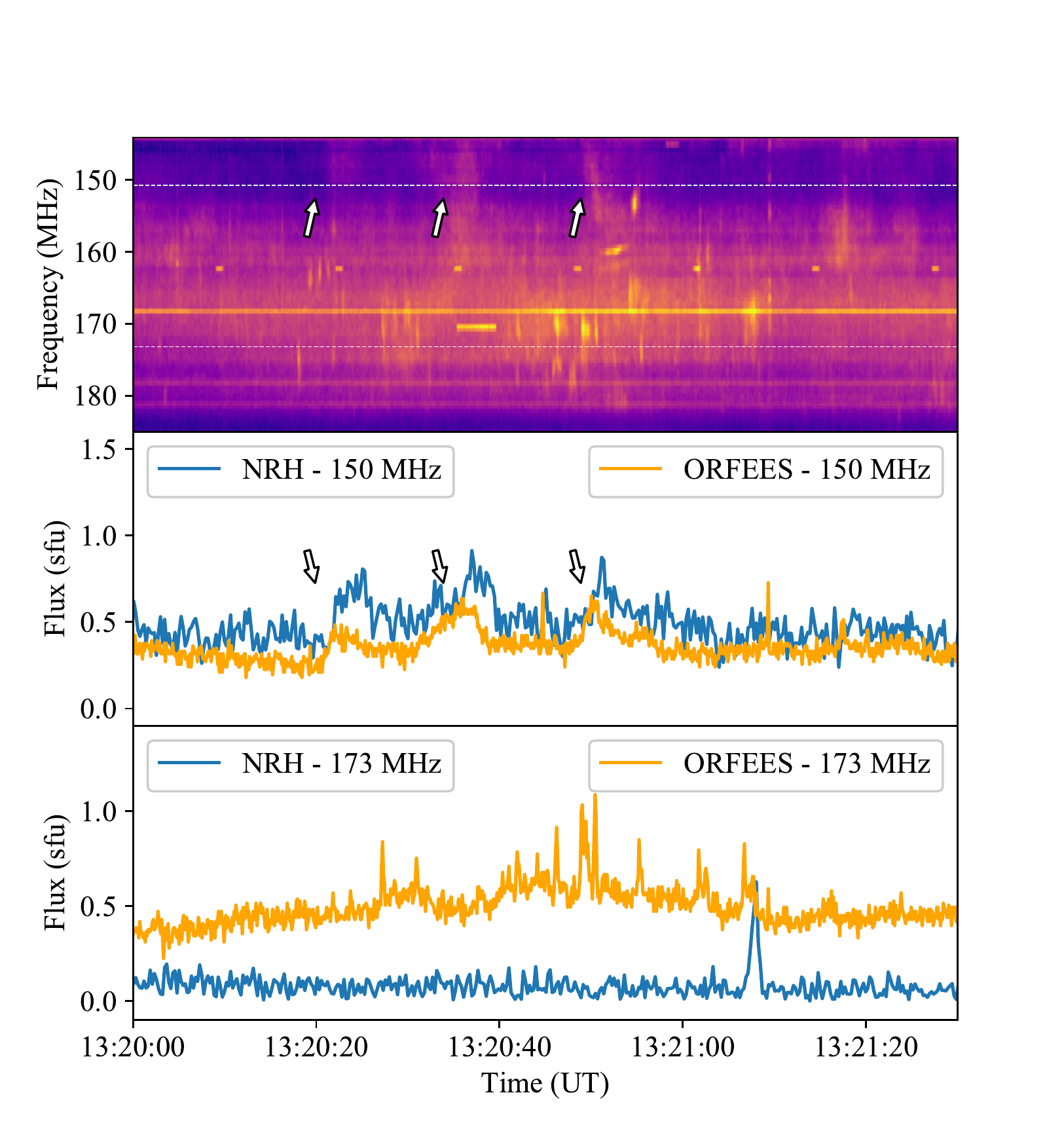}
    \caption{Spectral features corresponding to Source 3. (a) Zoomed-in view of the NDA dynamic spectrum showing a 90-s time period starting from 13:20:00~UT. (b) Time series of the NRH flux density of Source 3 in sfu and ORFEES normalised intensity in arbitrary units at 150.9~MHz. (c) Time series of the NRH flux density of Source 3 in sfu and ORFEES normalised intensity in arbitrary units at 150.9~MHz.}
    \label{fig:figa3}
\end{figure}

\end{appendix} 

\end{document}